\documentclass[10pt,conference]{IEEEtran}
\IEEEoverridecommandlockouts

\AtBeginDocument{%
  \providecommand\BibTeX{{%
    \normalfont B\kern-0.5em{\scshape i\kern-0.25em b}\kern-0.8em\TeX}}}

\usepackage{url}

\usepackage{breakurl}

\usepackage{graphicx}
\usepackage{color}
\usepackage{soul}

\usepackage{amsmath, amsthm}
\usepackage{tikz}
\usepackage{xspace}
\usepackage{xstring}
\usepackage{listings}
\usepackage{subfigure}
\usepackage{enumitem}

\usepackage{multirow}
\usepackage{booktabs}
\usepackage{comment}

\newcommand*\WC[1]{%
\begin{tikzpicture}[baseline=(C.base)]
\node[draw,circle,inner sep=0.2pt](C) {#1};
\end{tikzpicture}}

\newcommand*\BC[1]{%
\begin{tikzpicture}[baseline=(C.base)]
\node[draw,circle,fill=black,inner sep=0.2pt](C) {\textcolor{white}{#1}};
\end{tikzpicture}}

\newcommand{\ie}{\textit{i}.\textit{e}.}
\newcommand{\eg}{\textit{e}.\textit{g}.}
\newcommand{\PP}[1]{\vspace{2px} \noindent{\bf \IfEndWith{#1}{.}{#1}{#1.}}}
\newcommand{\cc}[1]{\mbox{\small \texttt{#1}}}

\newlist{questions}{enumerate}{1}
\setlist[questions]{label=\bf{RQ}}

\newcommand{\sys}{\mbox{\textsc{SmartMark}}\xspace}

\definecolor{darkpink}{rgb}{1,0.38,0.4}

\definecolor{codegreen}{rgb}{0,0.6,0}
\definecolor{codegray}{rgb}{0.5,0.5,0.5}
\definecolor{codepurple}{rgb}{0.58,0,0.82}
\definecolor{backcolour}{rgb}{0.95,0.95,0.92}

\lstdefinelanguage{JavaScript}{
  keywords={typeof, new, true, false, catch, function, return, null, catch, switch, var, if, in, while, do, else, case, break},
  keywordstyle=\color{blue}\bfseries,
  ndkeywords={class, export, boolean, throw, implements, import, this},
  ndkeywordstyle=\color{darkgray}\bfseries,
  identifierstyle=\color{black},
  sensitive=false,
  comment=[l]{//},
  morecomment=[s]{/*}{*/},
  commentstyle=\color{purple}\ttfamily,
  stringstyle=\color{red}\ttfamily,
  morestring=[b]',
  morestring=[b]"
}

\lstdefinestyle{solidity}{
	backgroundcolor=\color{backcolour},   
	commentstyle=\color{codegreen},
	keywordstyle=\color{magenta},
	numberstyle=\tiny\color{codegray},
	stringstyle=\color{codepurple},
	basicstyle=\ttfamily\footnotesize,
	language=C,
	basicstyle=\fontsize{7}{7}\ttfamily,
	breakatwhitespace=false,         
	breaklines=true,                 
	captionpos=b,                    
	keepspaces=true,                 
	numbers=left,                    
	numbersep=4pt,                  
	showspaces=false,                
	showstringspaces=false,
	showtabs=false,                  
	tabsize=2
}

\lstdefinestyle{solidityTextColor}{
	backgroundcolor=\color{backcolour},   
	commentstyle=\color{codegreen},
	keywordstyle=\color{magenta},
	numberstyle=\tiny\color{codegray},
	stringstyle=\color{codepurple},
	basicstyle=\ttfamily\footnotesize,
	language=C,
	basicstyle=\fontsize{7}{7}\ttfamily,
	breakatwhitespace=false,         
	breaklines=true,                 
	captionpos=b,                    
	keepspaces=true,                 
	numbers=left,                    
	numbersep=4pt,                  
	showspaces=false,                
	showstringspaces=false,
	showtabs=false,                  
	tabsize=2,
	moredelim=**[is][\color{red}]{@}{@}
}

\makeatletter
\newcommand{\linebreakand}{%
  \end{@IEEEauthorhalign}
  \hfill\mbox{}\par
  \mbox{}\hfill\begin{@IEEEauthorhalign}
}
\makeatother

\begin{document}

\title{\sys: Software Watermarking Scheme for Smart Contracts}

\author{\IEEEauthorblockN{Taeyoung Kim}
\IEEEauthorblockA{\textit{Department of Computer Science}\\
\textit{and Engineering}\\
\textit{Sungkyunkwan University}\\
Suwon, South Korea\\
tykim0402@skku.edu}
\and
\IEEEauthorblockN{Yunhee Jang}
\IEEEauthorblockA{\textit{Department of Electrical}\\
\textit{and Computer Engineering} \\
\textit{Sungkyunkwan University}\\
Suwon, South Korea \\
unijang@skku.edu}
\and
\IEEEauthorblockN{Chanjong Lee}
\IEEEauthorblockA{\textit{Department of Computer Science}\\
\textit{and Engineering}\\
\textit{Sungkyunkwan University}\\
Suwon, South Korea\\
leecj323@skku.edu}
\linebreakand
\IEEEauthorblockN{Hyungjoon Koo*}
\IEEEauthorblockA{\textit{Department of Computer Science}\\
\textit{and Engineering}\\
\textit{Sungkyunkwan University}\\
Suwon, South Korea\\
kevin.koo@skku.edu}
\and
\IEEEauthorblockN{Hyoungshick Kim*}
\IEEEauthorblockA{\textit{Department of Electrical}\\
\textit{and Computer Engineering} \\
\textit{Sungkyunkwan University}\\
Suwon, South Korea \\
hyoung@skku.edu}
\thanks{*Corresponding author.}
}

\maketitle

\begin{abstract}
A smart contract is a self-executing program on a blockchain to ensure an immutable and transparent agreement without the involvement of intermediaries. Despite its growing popularity for many blockchain platforms like Ethereum, no technical means is available even when a smart contract requires to be protected from being copied. One promising direction to claim a software ownership is software watermarking.
However, applying existing software watermarking techniques is challenging because of the unique properties of a smart contract, such as a code size constraint, non-free execution cost, and no support for dynamic allocation under a virtual machine environment.
This paper introduces a novel software watermarking scheme, dubbed \sys, aiming to protect the ownership of a smart contract against a pirate activity. \sys builds the control flow graph of a target contract runtime bytecode, and locates a collection of bytes that are randomly elected for representing a watermark.
We implement a full-fledged prototype for Ethereum, applying \sys to 27,824 unique smart contract bytecodes.
Our empirical results demonstrate that \sys can effectively embed a watermark into a smart contract and verify its presence, meeting the requirements of credibility and imperceptibility while incurring an acceptable performance degradation. %
Besides, our security analysis shows that \sys is resilient against viable watermarking corruption attacks; \eg, a large number of dummy opcodes are needed to disable a watermark effectively, resulting in producing an illegitimate smart contract clone that is not economical.
\end{abstract}

\begin{IEEEkeywords}
Smart contract, Software watermarking, Blockchain, 
Software copyrights
\end{IEEEkeywords}

\section{Introduction}
\label{intro}

Due to the advancements in blockchain technologies, blockchain-based 
smart contracts (hereinafter referred to as {\em smart contracts}) 
have received significant attention from both academia and industry 
over the last few years.
A vast number of smart contracts have already been deployed on blockchains 
(\eg, over 10 millions in 2020 on the Ethereum network~\cite{he2020characterizing}). 

As a smart contract has been adopted for business, we encounter a new (but familiar) challenge in protecting it when a smart contract owner needs to claim one's intellectual property right. Since a smart contract is a type of programming code, it is inherently prone to be plagiarized.
Although a smart contract is deployed in a binary form on a blockchain
like other software distributions, its size constraint (\eg, 24KB for Ethereum) makes reverse engineering relatively less painful with the state-of-the-art tools (\eg, Erays~\cite{Zhou18:Smartcontract}, Vandal~\cite{brent18:Smartcontract}, Gigahorse~\cite{Grech19:Smartcontract}).
Recent studies~\cite{pierro2021analysis, chen2021understanding} 
reveal that a vast amount of contract code blocks had been indeed cloned.
He et al.~\cite{he2020characterizing} identified 41 decentralized applications (DApps\footnote{A DApp is a collection of smart contracts incorporated with an interface on a website or an application, to interact with users.}) with 73 plagiarized DApps, which may cause a substantial financial loss to the original DApp creators. 
Besides, they showed that careless code reuse could bring unwanted results from a security perspective.

A well-known technique for protecting a software copyright
is software watermarking, a process of embedding a watermark 
$W$ into a program $P$ such that $W$ can be further detected or extracted to assert the ownership of $P$~\cite{collberg1999software}.
In essence, the underlying mechanism is that $W$ would be present 
when one copies $P$ even with the attempt of a corruption (\eg, modification). 
A plethora of software watermarking schemes~\cite{Peter1994, Keith1994, 
 davidson1996method, kang2021softmark, collberg1999software, jiang2009software, qu1998analysis, qu1999hiding, 
 balachandran2014function, monden2000practical, lu2014ropsteg,
 collberg2003sandmark, zuo2010zero, rovcek2021zero, wang2019ternary,
 ma2015software} have been introduced against a piracy.

However, applying the existing techniques to a smart contract is not viable due to its unique properties. First, hiding a watermark within a program would be difficult because the size of a smart contract is typically smaller than that of a conventional program. Consequently, it is possible for a skillful adversary to manipulate the watermark with static code analysis.
Second, a virtual machine such as Ethereum Virtual Machine (EVM) 
offers a different enviornment from a bare machine, making it infeasible 
to adopt prior approches like the return-oriented programming (ROP)-based 
scheme~\cite{ma2015software} or the function reodering scheme~\cite{kang2021softmark}.
Third, running a bytecode under EVM inevitably incurs
a transaction cost, which restricts any scheme that introduces 
additional code or data.
Fourth, EVM does not have a feature of dynamic memory allocation,
which disables the adoption of a prior scheme~\cite{collberg2004dynamic, 
collberg1999software} to utilize that feature.

Only a few attempts have been made to protect smart contract developers' intellectual property rights. 
Zhang et al.~\cite{Zhang20:Smartcontract} present code obfuscation techniques for a smart contract, especially written in Solidity~\cite{Ethereum2022Online}, which makes it difficult to decompile a smart contract. 
Yan et al.~\cite{Yan20:Smartcontract} propose a technique to increase the difficulty level of recovering a control flow graph from a smart contract bytecode by introducing four anti-reverse engineering code patterns. 
Although it is possible to raise the bar with the above techniques, they are still far from a comprehensive solution for verifying the originality of a smart contract (\ie, Has a contract been copied (partly or in full) from another?).

In this work, we present \sys, to the best of our knowledge, the first watermarking scheme on smart contracts that considers both varying requirements of a watermark (\ie, imperceptibility, spread, credibility, resiliency, capacity, efficiency) and unique properties of smart contracts (\ie, gas cost).
At a high level, {\sys} builds a control flow graph (CFG) from the runtime bytecode of a smart contract and randomly elects a series of bytes from the blocks selected across the CFG as a watermark.\footnote{A runtime bytecode is the execution body of a smart contract, and a creation bytecode consists of both a runtime and initialization bytecode  (\ie, constructor).} Next, {\sys} creates a data structure that holds essential information (\eg, the locations of elected bytes and the CFG generation method) to extract the watermark later. {\sys} privately keeps this data structure and only inserts the hash of the data structure into the creation bytecode of the smart contract for further verification of the watermark to claim the originality of the smart contract. Note that we assume that {\sys} does not add any extra code for watermarking to the runtime bytecode of the smart contract, indicating that the smart contract's functional behavior would remain the same without incurring additional gas costs to execute the smart contract.

The main benefits of our design choice are free from \WC{1}~an undesirable gas cost for a transaction as the size of a runtime bytecode stays intact, and \WC{2}~detection techniques based on a static analysis as no additional code is introduced for a watermark.
To this end, we develop a full-fledged prototype of \sys and demonstrate how our approach fulfills the requirements of a watermarking scheme on smart contracts at an acceptable cost.
Moreover, we collected all the blockchain blocks (about nine million blocks between 30 July 2015 and 21 June 2022) from the Ethereum Mainnet, and selected 27,824 unique bytecodes from those blocks to evaluate the effectiveness and efficiency of \sys.

The contribution of our paper is summarized as follows.
\begin{itemize}
    \item We present \sys, a novel software watermarking scheme that satisfies varying requirements of watermarking for smart contracts.
    \item We empirically evaluate \sys, demonstrating its effectiveness 
    and efficiency. 
    \item We thoroughly study 
    the resiliency of our watermarking scheme against viable attacks.
\end{itemize}

\section{Background} 
\label{background}
This section describes the background of the Ethereum smart contract and the virtual machine environment to run it.

\subsection{Smart Contract and License}
\PP{Smart Contract}
The term ``smart contract''~\cite{szabo_formalizing_1997} refers to a piece of programming code that is permanently stored and executed for processing transactions when predetermined conditions are met.
Smart contracts act as nodes or accounts on the blockchain by leveraging its tamper-resiliency, traceability, and transparency. %
Ethereum~\cite{wood2014ethereum} is one of the most popular and prominent blockchain-based smart contract platforms. The smart contracts on Ethereum are written in a high-level programming language such as Solidity. 
A Solidity code must be compiled into Ethereum bytecode~\cite{wood2014ethereum} to properly run on a blockchain, remaining immutable and indelible within a blockchain ledger. We design {\sys} by embedding a watermark into bytecodes (compiled from a smart contract written in Solidity at the Ethereum platform) and extracting it from the bytecodes. Therefore, {\sys} is agnostic to specific features of a high-level programming language.

\PP{License in Solidity} 
The Solidity compiler offers the means of a machine-readable SPDX license identifier~\cite{Ethereum2021Online} by default, which can be embedded into a bytecode as metadata (by inserting a specific license header into every source file). However, the license identifier differs from a digital watermark because it merely represents one of the standard licenses (\eg, MIT, Apache, BSD, Creative Commons) rather than specifying the actual ownership of a smart contract, being inappropriate for claiming a smart contract copyright.

\subsection{EVM and Bytecode}
\PP{Ethereum Virtual Machine (EVM)}
The Ethereum Virtual Machine (EVM) offers a stack-based runtime environment for smart contracts, where a chunk of bytecodes can be executed upon receiving a transaction. EVM maintains varying machine states that hold a data structure as an execution component, including accounts, balances, stack, memory, storage, and a program counter. EVM supports $150$ instructions~\cite{wackerow2021Online} where each comprises a single byte opcode (mnemonic) and zero or more operands.

\PP{Gas Cost} Ethereum introduces the notion of the execution fee, dubbed  \emph{gas}, for every EVM opcode (\eg, the opcode for multiplication consumes five units of gas) based on the computational and storage overheads of each opcode~\cite{wackerow2021Online}. This enables miners to obtain a reward for computational resources to store smart contracts and execute them. Besides, it can prevent denial-of-service (DoS) attacks that invoke a time-consuming function~\cite{2_Atzei}.

\PP{Bytecode for EVM}
Solidity emits two types of bytecode: \WC{1}~creation bytecode (\ie, \emph{init} bytecode) for initializing (\ie, constructor) and deploying a contract, and \WC{2}~runtime bytecode (\ie, \emph{deployment} bytecode) for executing the contract that is stored on a blockchain. The major difference in spending a cost is that a creation bytecode requires a gas once whereas a runtime bytecode consumes a gas in every transaction. Note that one can obtain runtime bytecodes on a blockchain and creation bytecode from a contract transaction log.%

\section{Smart Contract Watermarking}
\label{s:challenges}

This section describes the need for a watermark technique on smart contracts, challenges, varying requirements, and a threat model with viable attacks.

\subsection{Motivation and Challenges} 

\PP{Motivation}
Due to the nature of public blockchain platforms, even if the smart contract authors do not make source code publicly available, smart contracts can be exposed in a bytecode format and could be reused by anyone. Indeed, recent empirical studies~\cite{pierro2021analysis, chen2021understanding} reveal that code reuse in smart contracts is quite prevalent. For example, Chen et al.~\cite{chen2021understanding} discovered that $26\%$ of contract code blocks had been cloned ($14.6$ occurrences on average) from the $146K$ open-sourced projects, suggesting common patterns of code duplication in smart contracts. Although the power of reusability helps a smart-contracts-driven ecosystem to be rich, it may pose a severe threat to managing smart contracts' intellectual property rights (IPR). 
He et al.~\cite{he2020characterizing} demonstrated that over 96\% of 10 million contracts had duplicates. 
Besides, they reveal a case that entails
substantial financial losses (89,565.32 ETHs, about 30\% of the original market); 73 plagiarized DApps are copied from 41 original ones. Fomo3D is one of the popular DApps with over 10,000 active users and large transactions (40,000 ETHs) back in 2018, which has been victimized by numerous copycats.
Such a growing concern motivates our work for protecting the IPR of a smart contract. However, it is non-trivial to prevent reusing existing smart contracts. A possible solution would be to develop a software watermarking~\cite{collberg1999software} scheme to provide a technical means that claims the originality of a smart contract on demand: \eg, plagiarized DApps could have been disclosed with the scheme. Further, the presence of explicit watermarks helps in establishing ownership proof over legal disputes.

\PP{Challenges}
Applying prior software watermarking techniques to a smart contract is challenging due to the characteristics of a smart contract programming language. One of the biggest hurdles is that smart contracts typically have a small size (up to 24KB), making it difficult to conceal a watermark from the original smart contract code.
Another challenge is that running a bytecode under EVM comes with a (gas) cost (\ie, Ethereum transaction fee), possibly leading to avoiding any watermarking technique unless it stays the total cost intact.
Hence, it is evident that a watermarking scheme 
with a charge would not be welcomed 
even with the presence of the technique.
Lastly, the EVM environment for a smart contract does not allow for dynamically allocated memory, disabling the adoption of existing watermarking schemes~\cite{collberg2004dynamic, collberg1999software} that utilize dynamic allocation.

\PP{Goal} 
By nature, a blockchain is designed to confirm if a certain smart contract appears for the first time. 
However, a technical means is absent to verify that a suspicious smart contract is a replica of an existing smart contract (thereby violating IPR).
In this paper, we aim to provide
such a means to address this problem.
To the best of our knowledge, we introduce the first watermarking scheme for smart contracts that should be able to tackle the aforementioned challenges.
To demonstrate the effectiveness of \sys, our evaluation focuses on answering the following research questions (RQs):

\begin{itemize}
    {\item{\bf{RQ1: }}Is \sys able to protect smart contracts from being copied effectively and efficiently?}
    {\item{\bf{RQ2: }}Is \sys sufficiently resilient against a variety of adversarial attacks?}
\end{itemize}

\subsection{Requirements}
\label{ss:requirements}
Instead of reinventing the wheel for the requirements of a watermarking scheme, we adopt the general ones as with previous software watermark 
approaches~\cite{kang2021softmark, collberg1999software, myles2005evaluation, zeng2010robust, dalla2017software, dey2019software, qu1999hiding}:
imperceptibility, spread, credibility, resiliency, 
capacity, and efficiency. Besides, we define a cost (gas consumption) as another requirement for contract watermarking. %
\begin{itemize}[leftmargin=*]
\item \emph{Imperceptibility} ensures 
that a watermark must be scarcely perceptible,
that is, a smart contract with an embedded
watermark must be indistinguishable from 
the one without.
\item \emph{Spread} denotes how well
a watermark is distributed across the
whole smart contract code.
Typically, a well-scattered watermark
tends to be resilient against 
corruption attempts.
\item \emph{Credibility} ensures that
a watermark must be reliably verifiable, minimizing false positive or negative cases.
\item \emph{Resiliency} represents the robustness of a watermarking scheme against tampering attacks that aim to invalidate a watermark, including addition, subtraction, and distortion.
\item \emph{Capacity} represents the data rate of a watermark that can be encoded into a target contract. Considering a contract size constraint, the length of a watermark cannot exceed it.
\item \emph{Efficiency} represents a performance overhead (\ie, computational resource) that is needed for watermarking operations. We separately define a gas cost metric for smart contracts.
\item \emph{Cost} represents the amount of gas consumption for inserting and validating a watermark. We utilize a gas price per individual opcode as pre-defined in~\cite{wackerow2021Online}.
\end{itemize}
\begin{figure*}[h!]
    \centering
    \includegraphics[width=\linewidth]{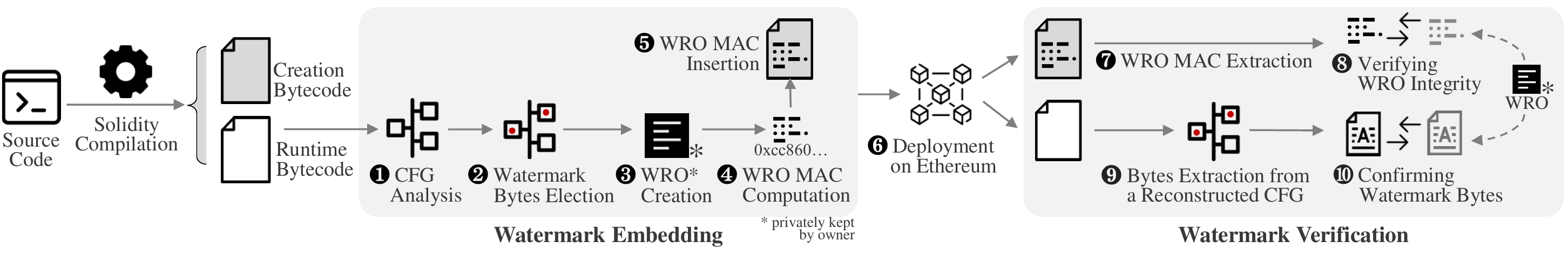}
    \caption{Overall \sys scheme for embedding and verifying
    a watermark for a smart contract. 
    We carefully elect bytes (\protect\BC{2}) that
    comprises a watermark 
    from the CFG  (\protect\BC{1}) 
    of a runtime bytecode, creating a 
    watermark reference object (WRO) (\protect\BC{3}).
    The original author secretly holds the WRO,
    and computes its hash (WRO MAC).
    The WRO MAC is computed and embedded to a creation bytecode (\protect\BC{4}, \protect\BC{5}), 
    followed by deploying it on the Ethereum 
    network (\protect\BC{6}).
    By verifying the extracted WRO MAC
    (\protect\BC{7}, \protect\BC{8}) from a creation bytecode with
    the extracted bytes from
    a reconstructed CFG (\protect\BC{9}),
    a watermark
    can be verified (\protect\BC{10}) on demand.
    }
    \label{fig:4_SmartMark}
\end{figure*}

\subsection{Threat Model} 
\label{ss:threatmodel}
The objective of our watermarking scheme for a smart contract is to thwart an adversary's considerable efforts with reasonable resources rather than suggesting an unbreakable scheme (as a fully motivated attacker could hardly be prevented).
With this in mind, in this paper, 
we assume a strong adversary
who is capable of 
\WC{1}~obtaining an open bytecode for a given smart contract,
\WC{2}~understanding our watermark scheme beforehand, and 
\WC{3}~performing arbitrary code manipulation
on a decompiled source code (Section~\ref{ss:resiliency}) or bytecode, 
attempting to tamper with a watermark.
We also assume that the adversary could collect as many smart contracts as possible
for further comparisons between them
(\ie, collusive attack).

\PP{Attack Types} 
We classify five viable attacks largely into two categories: passive attacks (denoted as ``P''), including unauthorized recognition and collusion (against imperceptibility), and active attacks (denoted as ``A''), including addition, deletion, and distortion (against credibility and resiliency). 
Our design principle does not necessarily conceal the presence of a watermark embedded in smart contracts at all. Hence, recognizing such information itself would not weaken the overall security of {\sys} unless an adversary could reveal the exact location of a watermark.

\begin{itemize}[leftmargin=*]
    \item \emph{(P) Unauthorized recognition} refers to an attack that identifies the location of a watermark in a target smart contract.
    \item \emph{(P) Collusion} refers to an attack that recognizes the location of a watermark in a target contract by comparison.
    \item \emph{(A) Addition} refers to an attack that embeds another watermark
    (\ie, adversary's ownership) into a target contract. %
    \item \emph{(A) Deletion} refers to an attack that eliminates a valid watermark from a target smart contract.
    \item \emph{(A) Distortion} refers to an attack that encompasses every transformation for damaging an existing watermark.
\end{itemize}

\section{\sys Design}
\label{design}

This section sketches the design of \sys that considers both the requirements for generic watermarks and the smart contract's idiosyncratic properties.
Notably, our scheme does not introduce additional runtime bytecode 
because implanting bytes would be revealed by a sophisticated attack as well as increasing an undesirable gas cost.

\subsection{Design Overview}

Fig.~\ref{fig:4_SmartMark} depicts
the overall process of \sys for embedding
and verifying a watermark for a smart contract.
First, the Solidity compiler generates both creation bytecodes
(\ie, contract constructor) and runtime bytecodes
(\ie, actual execution code under EVM).
For embedding a watermark, our scheme harnesses both creation and runtime bytecodes for watermarking over smart contracts.   
We first construct a CFG from a runtime bytecode (\BC{1} in Fig.~\ref{fig:4_SmartMark}), followed by electing a series of bytes that incorporates a watermark (\BC{2}). 
Then, we explicitly define a structure (dubbed \emph{watermark reference object}; WRO) that contains essential information to represent the watermark, which must be privately maintained by the owner of a smart contract (\BC{3}).
We compute the hash value of the WRO (\BC{4}) and insert it into a creation bytecode (\BC{5}) for further validating a watermark.
For brevity, hereinafter, we call such a hash value a \emph{WRO MAC} that represents a message authentication code for WRO. 
Once the creation bytecode is complete, it is deployed to the Ethereum network (\BC{6}).
It is worth noting that all bytecodes are publicly available in a tamper-proof fashion after deployment, which indicates that a WRO MAC is publicly accessible and immutable.
To verify the presence of a watermark in a target smart contract, a verifier (e.g., the owner of the original smart contract) uses the WRO containing the information about the watermark. For the validity of the WRO, the verifier first computes the hash value of the WRO and compares it with the WRO MAC extracted from a creation bytecode of the target contract (\BC{7}, \BC{8}). If the two values are matched, the WRO is valid.
Therefore, the verifier reconstructs a CFG from a target smart contract (\BC{9}), and confirms the presence of a watermark in the CFG with the information in the WRO (\BC{10}).

\subsection{Watermark Embedding for Smart Contracts}
\label{ss:embedding}
This section portrays \sys's watermark embedding for 
smart contracts under the hood.

\subsubsection{Design Choice}
\sys carefully considers the two types of bytecode with different properties; a runtime bytecode for electing the bytes that form a watermark and a creation bytecode for ensuring the integrity of a WRO. {\sys} does not introduce any extra code within a runtime bytecode since the execution of a smart contract inevitably requires a gas cost. Alternatively, we store a 32-byte WRO MAC as a variable in the creation bytecode because it only incurs a one-time cost during deployment.
Besides, we embed multiple ($N$) watermarks,
enabling a robust recovery in case of partial damage
(Section~\ref{ss:verification}).

\subsubsection{Strategic Byte Election}
\label{ss:strategy}
As illustrated in Fig.~\ref{fig:4_SmartMark},
we begin with constructing a CFG from a runtime bytecode.
However, not every byte accounts for a watermark in \sys because strategic insertion of a (cheap) dummy byte into every block within a target CFG can disturb watermark construction.
Therefore, we carry out two processes; 
choosing a set of candidate bytes that contribute to 
forming a watermark, and
randomly electing watermark bytes from the candidates. %

\begin{figure*}[h!]
    \centering
    \includegraphics[width=.81\linewidth]{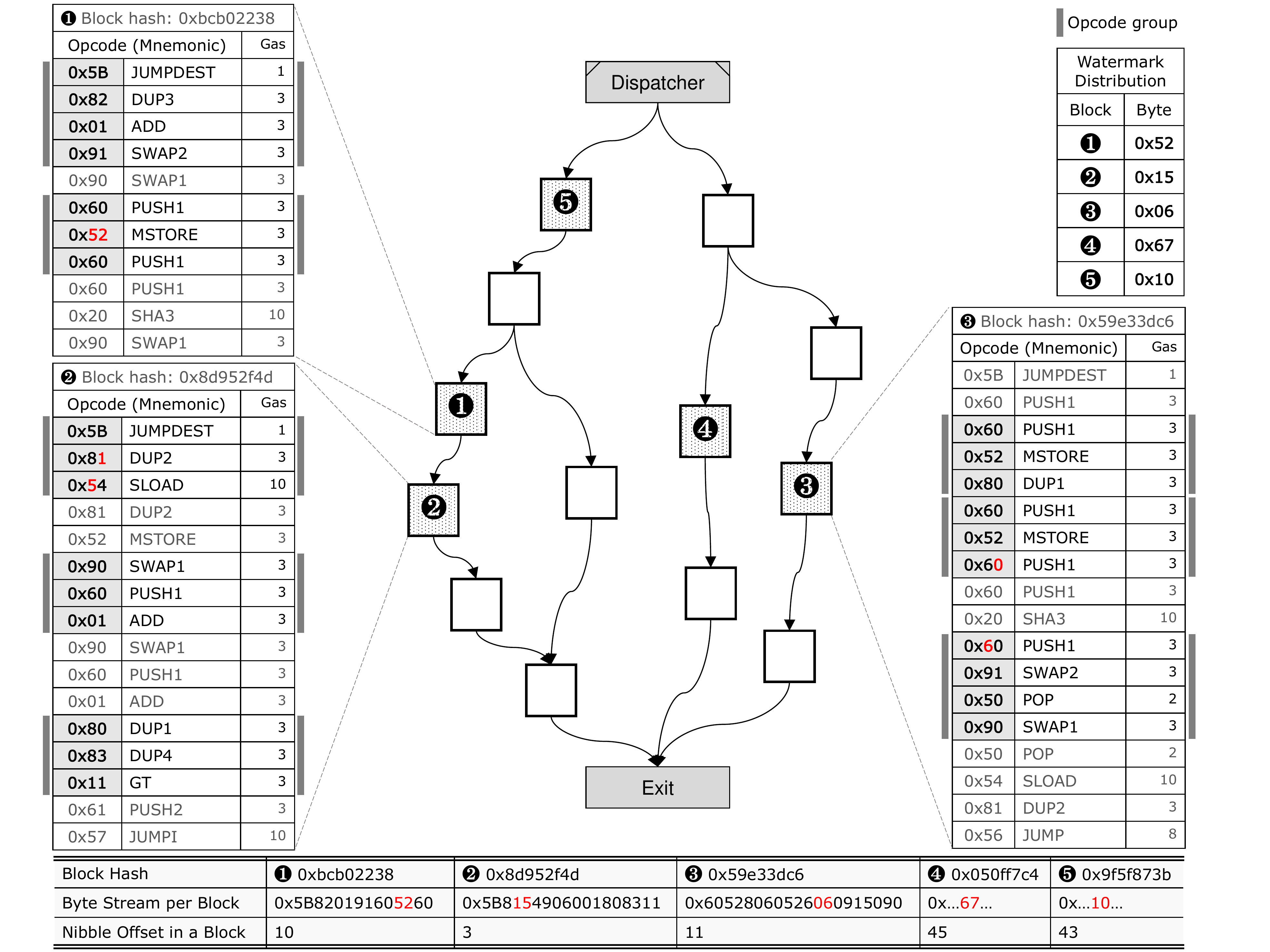}
    \caption{
    Example of strategic byte election
    with the \cc{ERC20} runtime bytecode
    for a five-byte watermark.
    Once CFG construction is completed, we randomly
    elect a series of opcode groups 
    (\ie, vertical lines in the block tables)
    from a watermarkble zone
    (Section~\ref{ss:embedding}), forming
    a final watermarkable byte stream.
    This example illustrates five blocks (\ie, dotted
    blocks; \protect\BC{1}-\protect\BC{5}) where three of them demonstrate
    how each byte of the watermark values (on the top right table) 
    has been distributed (\eg, red nibbles)
    across separate blocks. %
    The table on the bottom shows a block hash and
    a nibble offset in each byte stream for a block
    (value corresponding to the watermark),
    being recorded in a WRO.
    }
    \label{fig:4_CFG}
\end{figure*}

\PP{Watermarkable Zone}
We determine a code region that embraces a set of candidate bytes for a watermark, that is, a \emph{watermarkable zone} by excluding bytes unsuitable for watermarking.
First, we rule out a byte if it falls into a dispatcher function\footnote{The dispatcher is a built-in
function that points to user-defined (public) functions,
which should be invoked at the beginning
of runtime bytecode (\eg, the start function
in Fig.~\ref{fig:4_CFG}).} because it is commonly used for all smart contracts.
Second, we exclude all operands 
since they can be more fragile
to a watermark corruption
(\eg, modifying jumping destinations, 
immediate values on the stack).

\PP{Opcode Group}
From a collection of all bytes (\ie, list of opcodes) in
the above watermarkable zone, 
we group a sequence of opcodes together by
taking gas consumption into consideration.
An \emph{opcode group} can be determined with
three factors; \WC{1}~a cost threshold ($T$)
that represents the aggregate of gas costs
for consecutive opcodes,
\WC{2}~the size of a sliding window ($W$), and
\WC{3}~the maximum size of a group ($G$) or
the number of opcodes.
To exemplify, in Fig.~\ref{fig:4_CFG},
the first four consecutive bytes 
(\eg, \cc{0x5B, 0x82, 0x01, 0x91})
at the first block table 
(\BC{1}) can form a group
because the gas sum (1+3+3+3=10) exceeds
$T = 9$ in case of $G = 5$ and $W = 1$.
In the same vein, the next five consecutive bytes
(\eg, \cc{0x5B, 0x82, 0x01, 0x91, 0x90})
can hold another group as it meets
the requirements of the gas sum of 13
when the group size is 5.
Next, we move forward (\eg, starting from \cc{0x82})
to seek the next group candidates
until all blocks are covered.
The reasoning behind this process is that
we desire to generate as many opcode group
candidates (to choose from)
as possible while avoiding a fully overlapped group
under the distribution of opcode costs ($3$ as a mean value)\footnote{The most frequently appeared 
opcode is \text{PUSH1} in our dataset, whose cost is three. 
Note that more than three out of four
opcodes (76.2\%; 98,137,334 of 128,819,501)
consume a gas cost of three.},
the number of blocks per smart contract ($108.8$ on average),
and the number of opcode per block ($21.7$ on average).

\PP{Watermark Byte Election}
Concisely, once every opcode group in a watermarkable zone
is set up, we randomly subset all opcode groups,
followed by electing bytes for a watermark.
First, we randomly choose $R\%$ of all opcode groups.
The elected groups may be partially overlapped or consecutive,
forming a \emph{byte stream per block}.
Notably, the byte stream serves a basis
for further watermark verification.
Second, we elect $L$ bytes at random
so that each byte
can be present in a different block 
for a better spread
where $L$ is the number of distinguishing bytes
that a watermark demands.
As a concrete example, 
 Fig.~\ref{fig:4_CFG} illustrates 
the whole process of byte election with
the \cc{ERC20} smart contract~\cite{jdourlens2020Online}.
In this example, we show
three watermark bytes (\eg, \cc{0x52, 0x15, 0x06})
that are elected across eight opcode groups 
from three blocks
(\eg, 2, 3 and 3 per block) in case of $L$ = 5.
We adopt a unit of nibble ($4$ bits) 
to increase the likelihood of having the 
number of unique bytes from the byte stream
(one may want to use more fine-grained unit like a bit).
As an example, the second watermark byte of \cc{0x15}
in Fig.~\ref{fig:4_CFG} has been taken from
the second nibble at \cc{0x81} and 
the first nibble at \cc{0x54} (red letters).
Finally, a nibble offset is required
within a block for every elected watermark byte.
Then we bookkeep the information of 
a watermark including
opcode groups, block hashes, and nibble offsets
in a pre-defined structure 
(Figure~\ref{fig:4_WRO}).

\subsubsection{Watermark Reference Object (WRO)}
\label{ss:wro}
\begin{figure}[t]
    \centering
    \includegraphics[width=\linewidth]{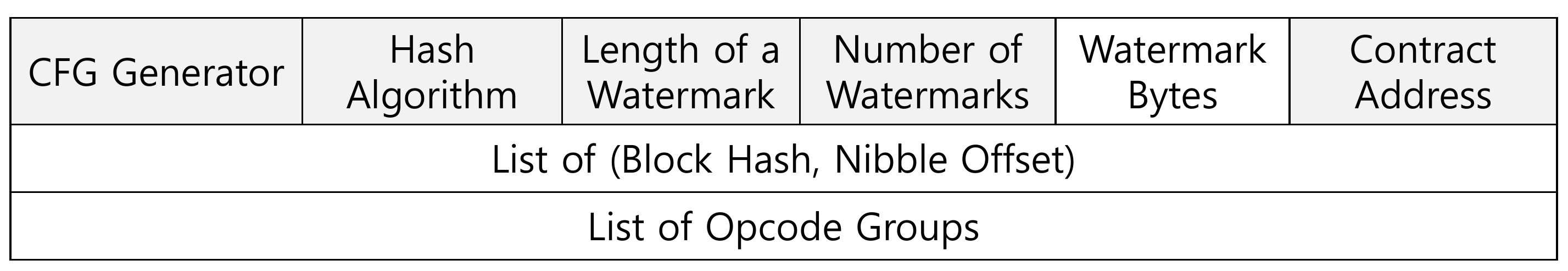} 
    \caption{Structure of a WRO. The grey fields represent fixed-length values while others are not.}
    \label{fig:4_WRO}
\end{figure}

We define a structure, dubbed WRO, which records crucial information to validate a smart contract watermark, including block identifiers, nibble offsets within each block, opcode groups, the length of a watermark as well as the watermark itself.
Note that a block identifier is a hash value of a byte stream
(\ie, subset of all opcodes), not of an entire block.
Similarly, an offset represents
a distance within a byte stream 
for better resiliency.
Fig.~\ref{fig:4_WRO} shows the structure of the object in detail.
Additionally, it contains a CFG generation tool identifier\footnote{\sys predominantly relies on deterministic CFG construction from a CFG generator (e.g., EtherSolve), regardless of its correctness. Hence, verifying the presence of a watermark should be feasible as long as extracting the identical CFG from a certain contract. In this regard, a WRO contains a CFG tool identifier, including a specific version.}, a block hash algorithm (each block takes the first four bytes of the byte stream's hash for identification), the number of
a watermark (multiple watermarks are embedded for robustness against partial corruptions), and a contract address. 
As presented in Section~{\ref{ss:embedding}}, we compute the hash of the WRO (\ie, WRO MAC), and store it to a variable in a constructor (Listing~\ref{list:4_constructor}).
\lstset{style=solidity, label=list:4_constructor, frame=tb,
	caption=Example of inserting a WRO MAC in Solidity. 
	A constructor holds a variable for the MAC
	that resides in a creation bytecode 
	after deployment.}
\begin{lstlisting}
contract Watermark {
    bytes WRO_MAC;
    constructor() {
        // WRO MAC with Keccak-256
        WRO_MAC = "cc860417...fc6b";
    }
}
\end{lstlisting}

\subsection{Watermark Verification}
\label{ss:verification}

A watermark verification entails three main phases as follow.
First, given a WRO, we can compute the hash of the WRO and compare it with the stored WRO MAC extracted from the creation bytecode to ensure the integrity of the WRO.
Second, given a target runtime bytecode, 
watermark verification reconstructs
a CFG with the same CFG generator during an embedding process,
followed by creating a byte stream (per block)  with a list of opcode groups in the WRO.
The way to seek a certain byte stream in a CFG is through a series of byte-level searches at every target block.
Third, a verifier can compute every block hash
with that byte stream, and find watermark bytes from each block with nibble offset information. The verifier can successfully recover a watermark unless an adversary corrupts one of the byte streams contributing to the watermark. Note that multiple watermarks are inserted to enhance \sys's resilience against corruptions.

\section{Implementation}
Our \sys prototype is written in Python 3.9. We leverage EtherSolve~\cite{4_EtherSolve} into disassembling a smart contract bytecode and generating a CFG. If the CFG holds multiple blocks having the same bytecode, we consider only one block so that all blocks are uniquely different for \sys.

\PP{Hash Algorithm}
The case that \sys employs a hash algorithm
is twofold (Section~\ref{ss:embedding}): one for generating the hashes of all blocks in a CFG, and one for creating a WRO MAC.
In both cases, we utilize the Keccak-256 hash algorithm
whose output is a fixed length of $256$ bits.
Note that we take the first four bytes 
of a Keccak-256 digest to represent a block hash,
which is equivalent to generating an identifier 
for a user-defined function when compiling a smart contract into bytecode in EVM. 

\PP{Hyperparameters}
We deliberately leave a handful of hyperparameters
to be able to be adjusted (Section~\ref{ss:embedding}) 
in need for \sys.
For determining opcode groups, we introduce a gas threshold ($T$) with the size of a sliding window ($W$) and that of an opcode group ($G$), which assists in electing bytes to meet the requirements (Section~\ref{ss:requirements}). %
The ratio of elected opcode groups is set to $R\%$, which forms a byte stream per block.
\
The length of a watermark is set to $L$ bytes, and
the current \sys implementation elects $L$ blocks
accordingly (\ie, electing a single byte per block).
Lastly, the number of embedded watermarks ($N$) is  
for robust verification where
multiple watermarks may hold different values.
It is possible to insert a different length for each watermark, however, we use the same lengths for a straightforward security analysis.
The following enumerates concrete hyperparameters
in our experiment for \sys:
$T = 9$, $W = 1$, $G = 5$, $R = 0.2$, 
$N$ = $\{1, 3, 5, 7\}$, 
and $L$ = $\{10, 20\}$.
In general, care must be taken in setting parameters
as follows:
\WC{1}~$N$ and $L$ rely on the number of blocks 
in a smart contract,
\WC{2}~$R$ strikes a balance between the probability
of successful attacks (\eg, too high $R$ may increase
a distortion attack with a higher chance of choosing
precise opcode groups)
and the number of opcode group candidates
available (\eg, too low $R$ may fail to 
have sufficient candidates), and
\WC{3}~we empirically advise $T = 9$ and $G = 5$
for both effectiveness and efficiency.

\section{Evaluation} 
\label{evaluation}
In this section, we present how well \sys meets the requirements of a smart contract watermark (Section~\ref{ss:requirements} with \textbf{RQ1}), and analyze its robustness from a security perspective (Section~\ref{ss:threatmodel} with \textbf{RQ2}).
We evaluate \sys{} on a 64-bit Ubuntu 18.04 system 
equipped with Intel(R) Xeon(R) Gold 6230 2.10 GHz 
and 567GB RAM.

\PP{Dataset}
We collected all fifteen
million blocks (mined during the period 30 July 2015 to 21 June 2022) from the Ethereum Mainnet, which incorporates $4,112,336$ smart contracts.
Then, we obtain $445,930$ unique runtime bytecodes 
($10.8\%$ of the whole) after eliminating 
all byte-level equivalent ones.
We leverage EtherSolve~\cite{4_EtherSolve}
to obtain CFGs for $284,344$ contracts, followed by
taking $178,119$ contracts ($62.6\%$) that
contain a unique watermarkable zone
(recall that operands and dispatcher blocks
are excluded for our scheme).
We indeed observe a high ratio of code duplication
in smart contracts, aligned with the
previous findings~\cite{chen2021understanding}.
We manually confirm that a considerable number of
code reuse cases arise from the 
standards~\cite{ethereum-standard} introduced by 
the Ethereum community such as 
ERC-20 (standard interface for fungible tokens) and
ERC-721 (standard interface for non-fungible tokens).
In this respect, we group analogous contracts together
to demonstrate the effectiveness of our watermarking scheme.
We perform DBSCAN
clustering~\cite{6_clustering} with the similarity metric
of $\text{max}(\frac{len(a \cap b)}{len(a)},\frac{len(a \cap b)}{len(b)}$) where $a$ and $b$ denote the CFG blocks 
for the two contracts, $A$ and $B$.
This is because \sys mainly targets 
disparate bytecodes (\ie, program logic), 
excluding a common component 
like ERC20 that causes an overlapping between contracts.
We finally obtain $27,824$ smart contracts
out of 178K ($15.6\%$)
distinguishing sample contracts for our experiment.
The average size of the smart contracts in 
our dataset is $5.8KB$ with the standard deviation
of $4.6KB$ (median: $4.5$ KB).

\subsection{Imperceptibility}
\label{ss:imperceptibility}
In \sys, a watermark is completely imperceptible as long as its WRO is privately kept by the smart contract owner. To embed a watermark into a smart contract, {\sys} does not add any additional code and data except a WRO MAC on its creation bytecode; the watermark is constructed with watermark bytes randomly elected from its runtime bytecode, resulting in that watermark bytes are indistinguishable from the other bytes in the runtime bytecode. Perhaps, a sophisticated adversary can identify the WRO MAC from the creation bytecode. However, we note that a WRO MAC is a cryptographic hash value that is irreversible. Therefore, one cannot obtain fruitful information about WRO from the WRO MAC.

\subsection{Spread}
\label{ss:spread}
Going back to Fig.~\ref{fig:4_CFG}, \sys
intentionally picks a single byte
from a byte stream per block,
resulting in a well-scattered watermark
across different blocks in a CFG.
In this example, five out of
$13$ blocks (around $38\%$) are covered
for a watermark, and a lengthier one
would be more dispersed.
Besides, by design, \sys allows
one to insert multiple watermarks so that
it could increase verifiability
even with a single watermark being survived.
Such a design choice makes \sys robust
against varying attacks by lowering
the possibility of damaging every
watermark simultaneously.
Oftentimes, it would be excessively costly for an adversary
to disrupt a well-distributed watermark, hampering
further transactions (Section~\ref{ss:resiliency}).

\subsection{Capacity}
\label{ss:capacity}
In practice, the total length (\ie, capacity) of a watermark
(or multiple watermarks) is bounded
by the number of blocks in a CFG because
our scheme relies on how to choose
opcode groups in a watermarkable zone.
In our experimental setting,
we elect one byte from a single block.
In general, the capacity of our watermarking scheme
is determined by the number of watermarkable blocks available
in a smart contract, rather than the size of the contract.

\subsection{Efficiency}

To demonstrate the efficiency of \sys,
we run experiments of
embedding and verification processes 10 times to measure CPU time.
Note that we exclude any smart contract
that does not conform to the code size limit
(\ie, EIP-170~\cite{eip170}) in our experiments.
Furthermore, we measure sub-phases
for watermarking operations (Table~\ref{table:5_Each_Phase}); 
the phase of electing watermark bytes dominates
the entire embedding time ($96.7\%$) whereas that of
creating (watermarkable) bytestream does the verification time ($70.1\%$).
A wide range of variations mainly arise from
processing time that largely depends on
the number of blocks pertaining to a watermark,
rather than the size of a smart contract.
We observe that a verification process (mean: $17,258$ milliseconds)
takes approximately 1.5 times longer
than an embedding process (mean: $11,088$ milliseconds),
but is still overall acceptable in practice
(\eg, within $20$ seconds).

\begin{table}[t]
    \caption{Breakdown of embedding and
        verification time on average. 
        Electing bytes for a watermark and
        creating a bytestream dominate
        embedding and verification time,
        respectively.
        }
    \begin{center}
    \resizebox{0.98\linewidth}{!}{

        \begin{tabular}{l|l|r|r}
        \hline
        \textbf{Process} & \textbf{Phase} & \textbf{Ratio (\%)} & \textbf{Time (ms)} \\
        \hline
        \multirow{3}{*}{\textbf{Embedding}} 
        & Opcode grouping 
        & 3.38 & $363.88 \pm 410.62$ \\
        & Watermark byte election & 96.69 & $10,720.59 \pm 13,083.91$ \\
        & WRO creation & 0.03 & $3.38 \pm 2.49$\\
        \hline
        \multirow{3}{*}{\textbf{Verification}} 
        & WRO verification & 0.01 & $0.71 \pm 0.70$\\
        & Bytestream creation & 70.06 & $12,089.79 \pm 14,421.60$\\
        & Hash discovery & 29.93 & $5,165.84 \pm 5,097.88$\\
        \hline
        \end{tabular}
    }
    \end{center}
    \label{table:5_Each_Phase}
\end{table}

\subsection{Cost}
Recall that \sys does not introduce any
additional routine on a runtime bytecode,
staying the original execution gas intact.
Instead, \sys increases a (one-time) creation cost
due to a WRO MAC in a constructor (Listing~\ref{list:4_constructor} in Section~\ref{ss:embedding})
where its cost is closely proportional to the size of the WRO MAC.
In case of embedding a 256-bit WRO MAC using
Keccak-256 in our implementation, an additional
cost ranges from 48,540 to 53,100 gas.
Such cost variation mostly arises from varying opcodes 
depending on the original context of the constructor 
when inserting a WRO MAC.
Although a gas price frequently fluctuates, as of writing,
the additional gas consumption is around $4.27\sim4.67$
US dollars (USD) with the exchange rate of $90\sim100$ gas
per bit. %

\subsection{Credibility}
\label{ss:Credibility}

\begin{table}[t]
	\centering
	\caption{
	Ratio of unique watermarks 
	by the whole size ($L \times N$) of an embedded watermark.
	$L$ and $N$ denote the length of a watermark ($L$) 
	and the number ($N$) of watermarks, respectively.
	A watermarkable contract represents
	the contract that incorporates 
	a sufficient number of watermarkable zones.
	}
	\resizebox{0.85\linewidth}{!}{
		\begin{tabular}{ll|r|rr}
		\toprule
		\multirow{2}{*}{\textbf{L}} & \multirow{2}{*}{\textbf{N}} & \textbf{Watermark} & \textbf{\# Unique Watermarks} & \multirow{2}{*}{\textbf{Ratio}} \\ 
		& & \textbf{Size (bytes)} & \textbf{(Watermarkable Contracts)} & \\
		\midrule
		\multirow{4}{*}{10} & 1 & 10 & 27,704  (27,823) & 99.995\% \\ 
		& 3 & 30 & 23,062    (23,378) & 99.986\% \\
		& 5 & 50 & 18,098    (18,511) & 99.978\% \\
		& 7 & 70 & 15,140    (15,620) & 99.970\% \\ \midrule
		\multirow{4}{*}{15} & 1 & 15 & 27,764  (27,764) & 100.000\% \\ 
		& 3 & 45 & 19,482    (19,485) & 99.999\% \\
		& 5 & 75 & 15,144    (15,156) & 99.999\% \\
		& 7 & 105 & 13,009   (13,030) & 99.998\% \\ \midrule
		\multirow{4}{*}{20} & 1 & 20 & 26,679  (26,679) & 100.000\% \\ 
		& 3 & 60 & 16,874    (16,874) & 100.000\% \\
		& 5 & 100 & 13,311   (13,311) & 100.000\% \\
		& 7 & 140 & 11,646   (11,646) & 100.000\% \\
	    \bottomrule
		\end{tabular}
	}
	\label{table:5_FP}
\end{table}
Credibility is one of the essential requirements for a watermarking scheme, which ensures that it can be reliably extracted for proof of ownership. 
We assess \sys by confirming that a watermark from WRO must be unique, that is, the watermark cannot be present elsewhere
but the original contract.
Table~\ref{table:5_FP} summarizes the total bytes of the watermark(s) ($L \times N$) with the length ($L$) and the number ($N$) of the watermark(s) inserted into a smart contract, and the ratio of contracts that hold unique watermarks accordingly.
Due to the constraint of an embedding watermark size that relies on a smart contract size, we compute the ratio of uniqueness based on the number of watermarkable contracts where a certain size (\ie, $L \times N$ bytes) of a watermark can be embedded.
For example, a single 10-byte long watermark can possibly be inserted into $27,823$ ($99.999\%$) out of $27,824$ contracts in total, resulting in $27,704$ contracts holding unique values ($99.995\%$).
Empirically, the uniqueness ratio is proportional to the length of a watermark whereas inversely proportional to the number of the watermark. 
Hence, we advise striking a balance between the length and the number for fulfilling both capacity and spread properties (as well as credibility).

\subsection{Resiliency}
\label{ss:resiliency}

In this section, we show how \sys can defend against the attacks presented in Section~\ref{ss:threatmodel}. Note that Section~\ref{ss:imperceptibility} covers \emph{unauthorized recognition}.

\subsubsection{Collusion} 
A collusive attack would help identify a smart contract's WRO MAC by analyzing the differences between the creation bytecodes of several smart contracts because all WRO MAC values always have the same fixed length, \eg, SHA-256 produces a 32 bytes hash value, and have a higher entropy than the other variable values. As mentioned in Section~{\ref{ss:imperceptibility}}, however, we note that the presence of a WRO MAC itself does not practically help recognize its corresponding watermark because a WRO MAC is a cryptographic hash value that is irreversible.

\subsubsection{Addition} 
We note that an adversary can embed a watermark $W_{a}$ of one's choice into a smart contract using {\sys} in the same manner even when a watermark $W_{o}$ is already present in a smart contract. In such a case, one can verify the presence of both watermarks ($W_{a}$ and $W_{o}$) with the original smart contract owner's own WRO and the adversary's own WRO, respectively. However, it is easily recognizable that $W_{o}$ was embedded prior to $W_{a}$ because transactions are permanently recorded on the Ethereum blockchain network in chronological order; checking if the WRO MAC of an early-deployed watermark will do. %

\subsubsection{Deletion} 
With our scheme, a watermark is constructed with existing opcodes in an original smart contract. %
In theory, a watermark deletion may be possible when
an adversary could replace one or more opcodes used for the watermark with others, however, such
semantic-preserving code transformation that
maintains a reasonable cost would be quite challenging.

\subsubsection{Distortion}
\label{sss:6_distortive}
One of plausible and powerful attacks to corrupt
our watermark scheme is a distortion attack
with arbitrary transformations.
We randomly choose 2,500 smart contracts whose
source code have been publicly available.
In this experiment, we set
the length of a watermark ($L$) to be 15,
and the number of watermarks ($N$) to be 3.

\PP{Empirical Results} 
We conduct various distortion experiments
for thwarting an embedded watermark
with the following five different types of transformations
as suggested by Chen et al.~\cite{chen2021understanding}:
\WC{1}~adjusting a function visibility (\eg, access modifier alters \cc{public} to \cc{private}, or the other way around), \WC{2}~updating an inheritance relationship (\eg,  arbitrary subcontract is added and inherited), \WC{3}~introducing an additional state variable (\eg, original state variable refers to an arbitrary state variable), \WC{4}~defining a new event and function (\eg, additional function has been added to an original contract), and \WC{5}~adding a statement (\eg, arbitrary statement is added to an original function).
Table~\ref{tab:credibility_exp02} summarizes the 
results of the above distortion attempts against
our scheme.
\sys shows the robustness of an individual attack
(\ie, 99\% or above), and applying all transformations
barely drops the ratio of appropriate verification 
(\ie, $98.9\%$).

\begin{table}[t]
	\centering
	\caption{
	Experimental results of varying distortion attacks against an embedded watermark on 2,500 smart contracts, which 
	shows the robustness of the \sys scheme.
	}
	\resizebox{0.98\linewidth}{!}{
		\begin{tabular}{l|r|rr}
		\toprule
		\textbf{Transformation Type (Distortion)} & \textbf{\# Verified Contracts} & \textbf{Ratio} \\
		\midrule
		{\WC{1}} Adjusting a function visibility & 2,490 & 99.60\% \\
		{\WC{2}} Updating an inheritance & 2,495 & 99.80\% \\
		{\WC{3}} Introducing an additional state variable & 2,489 & 99.56\% \\
		{\WC{4}} Defining an event \& function & 2,500 & 100.00\% \\
		{\WC{5}} Adding a statement & 2,475 & 99.00\% \\
		{\WC{+}} Applying all the above & 2,472 & 98.88\% \\
	    \bottomrule
		\end{tabular}
	}
	\label{tab:credibility_exp02}
\end{table}

\PP{Theoretical Analysis}
As presented in Section~{\ref{ss:imperceptibility}}, an adversary cannot distinguish watermark bytes from other bytes in a smart contract.
Therefore, for each block in the CFG of a target contract containing a watermark byte, the best distortion strategy at the bytecode level would be guessing an opcode group used for the watermark and adding it to the block.
If the adversary's guess is correct, the newly added opcode group would be used to form a watermarkable byte stream with the existing opcode groups for the watermark to compute the CFG block hash.
Consequently, the valid block hash contained in the WRO would not match the new block hash, leading a watermark corruption.

\begin{figure}[t]
    \centering
    \includegraphics[width=0.7\linewidth]{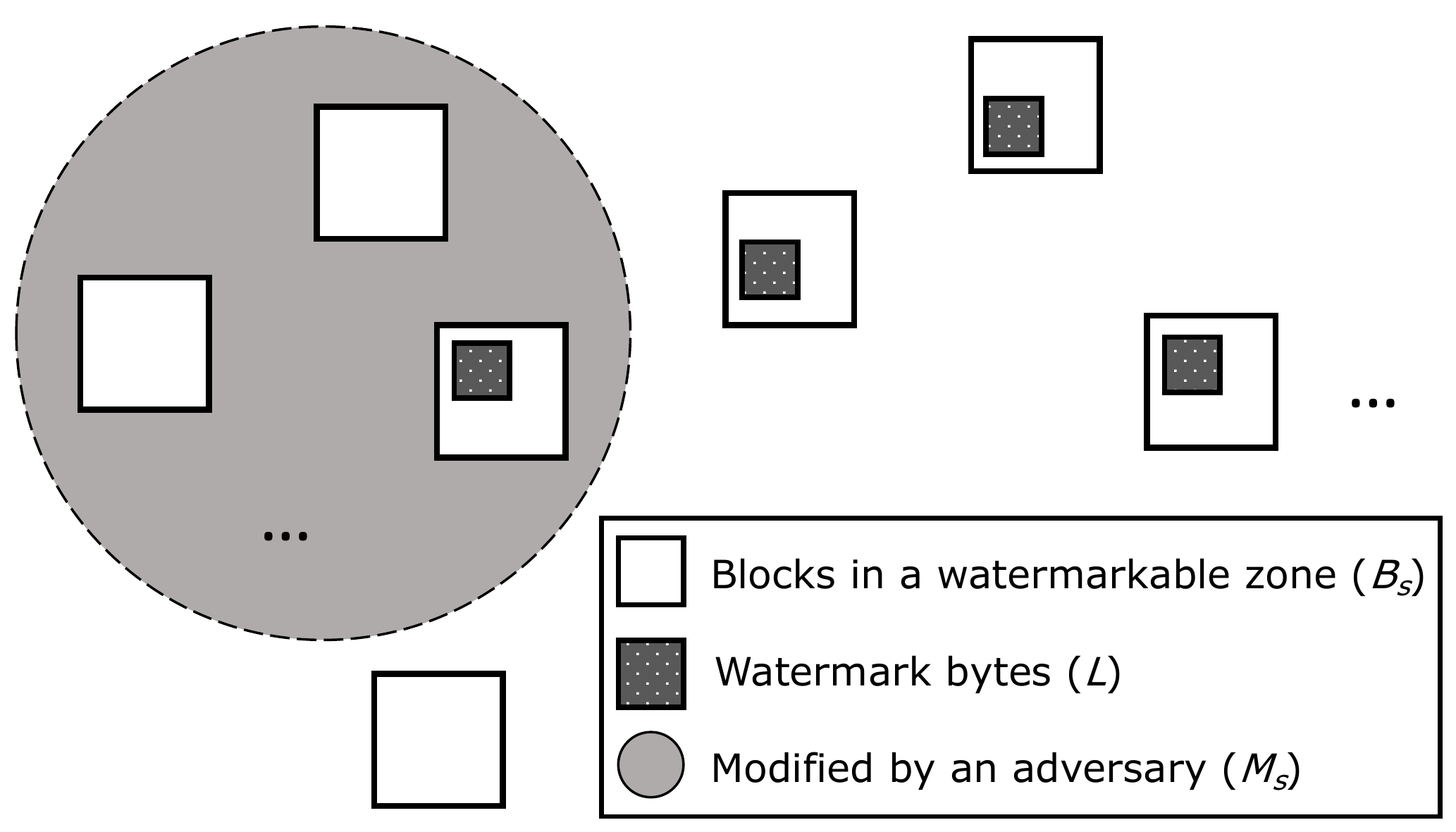}
    \caption{Illustrative parameters to compute 
    the probability of an attack success,
    $P_{attack}(L, B_s, M_s)$. 
    An empty rectangle represents a candidate block
    in a watermarkable zone, where the rectangle with
    a dark box contains a watermark byte.
    The dotted circle represents an area that
    includes the candidate block(s) modified by an adversary.
    }
    \label{fig:Visual illustration of the parameters}
\end{figure}

Given a smart contract of $s$, the probability of an attack success ($P_{attack}$ $(L, B_s, M_s)$) with a distortion can be computed by Equation~\eqref{eq: distortive1} where $L$, $B_s$, and $M_s$ represent the length of the watermark, the number of candidate blocks (in a watermarkable zone) for $s$, and the number of candidate blocks (in a watermarkable zone) that has been modified by an adversary against $s$, respectively. Figure~\ref{fig:Visual illustration of the parameters} concisely illustrates those parameters. Watermark bytes are scattered in candidate blocks in a watermarkable zone. Suppose that an adversary was able to modify the block(s) of one's choice where some of which could contain a watermark byte. %

\vspace{-0.3cm}
\begin{center}
\begin{equation}
P_{attack}(L, B_s, M_s) =\frac{\sum_{i=1}^{min(L, M_s)}\binom{B_s}{M_s}\binom{M_s}{i}\binom{B_s-M_s}{L-i}}{\binom{B_s}{L}\binom{B_s}{M_s}}
\label{eq: distortive1}
\end{equation}
\end{center}

In Equation~\eqref{eq: distortive1}, the denominator represents the number of all possible ways by choosing \WC{1}~$L$ blocks from $B_s$ blocks disallowing duplicates, and \WC{2}~$M_s$ blocks from $B_s$ blocks disallowing duplicates, respectively. The numerator represents the number of successful distortion attacks. When an adversary modifies $M_s$ blocks from $B_s$ blocks, the distortion attack would be successfully performed if those modified blocks contain at least one block that contains watermark bytes. In Equation~\eqref{eq: distortive1}, $i$ represents the number of blocks that contains actual watermark bytes out of $M_s$ blocks modified by an adversary. In this scenario, the number of successful distortion attacks can be interpreted as the number of possible ways to choose $M_s$ blocks from $B_s$ blocks, $i$ blocks from those $M_s$ blocks, and $L-i$ blocks from the remaining $B_s-M_s$ blocks disallowing duplicates. Finally, $\binom{B_s}{M_s}$ can be canceled out in the numerator and the denominator.

To enhance the resiliency of \sys against distortion attacks, we can insert $N$ watermarks that do not overlap each other where $N > 1$. In this case, an adversary needs to corrupt all $N$ watermarks for a successful attack. We presume that all watermarks have the same length for simplicity of analysis. With the notation of $P_{attack}(L, B_s, M_s, N)$, the probability of an attack success with $N$ watermarks on a smart contract $s$. $P_{attack}(L, B_s, M_s, N)$ can be simply expanded from $P_{attack}(L, B_s, M_s)$ as in Equation~\eqref{eq: distortive5}.
\vspace{-0.3cm}
\begin{center}
\begin{equation}
P_{attack}(L, B_s, M_s, N) = P_{attack}(L, B_s, M_s)^N
\label{eq: distortive5}
\end{equation}
\end{center}

We compute $P_{attack}(L, B_s, M_s, N)$ for each smart contract in the 27,824 smart contracts with varying parameters such as $L$, $N$, and $\alpha$ where $\alpha$ represents the ratio of the maximum allowable gas cost for the opcodes added by an adversary (hereafter referred to as ``attack budget'') to the total gas cost to execute a target smart contract.  
To compute $P_{attack}(L, B_s, M_s, N)$, we need to concretely obtain $B_s$ and $M_s$ from the smart contract $s$. With the hyperparameters of $T = 9$, $W = 1$, $G = 5$, and $R = 0.2$, as watermarkable blocks for $s$ are determined by our \sys implementation, we can empirically obtain $B_s$. 
However, $M_s$ cannot be determined by \sys because $M_s$ depends on an adversary's choice -- an adversary needs to correctly guess an opcode group for a candidate block containing watermark bytes and add it to the block. Therefore, given a smart contract $s$, we compute the expected value of $M_s$ ($E(M_s)$) with specific parameter values for $s$. The attack budget can be computed as $\alpha \cdot \Psi_s$ where $\Psi_s$ represents the total gas cost to execute the smart contract $s$. For example, when $\Psi_s = 1,000$ and $\alpha = 0.5$, the attack budget would be 500. In other words, 50\% of the gas cost ($\alpha=0.5$) is more needed to run a clone DApp. From the attack budget $\alpha \cdot \Psi_s$, we can compute the maximum number of opcode groups added to $s$ for a distortion attack as $\lfloor\alpha \cdot \Psi_s / T\rfloor$ because the gas cost for each opcode group, which corrupts a watermark byte effectively, is greater than or equal to $T$. In the previous example, when $T=9$, the maximum number of opcode groups added to $s$ is 55. We assume that an adversary knows $T$ used for \sys for simplicity of analysis. Then, the adversary's best attack strategy is to select $\lfloor\alpha \cdot \Psi_s / T\rfloor$ blocks out of $C_s$ blocks and add an opcode group to each block one by one where $C_s$ represents the number of the blocks that contains opcode groups, which requires a gas cost greater than or equal to $T$. With this strategy, the probability of selecting a candidate block in a watermarkable zone is $B_s / C_s$ whenever an adversary modifies a block. Thus, $E(M_s)$ can be computed by multiplying $\lfloor\alpha \cdot \Psi_s / T\rfloor$ by $B_s / C_s$ as follow:

\vspace{-0.1cm}
\begin{equation}
\begin{aligned}
E(M_s) = \left\lfloor\frac{\alpha \cdot \Psi_s}{T}\right\rfloor \cdot \frac{B_s}{C_s}
\label{eq: distortive3}
\end{aligned}
\end{equation}

\begin{figure}[t]
     \centering
     \subfigure[$\alpha=0.2$ with $L=10$ (left) and $L=20$ (right)]{
         \includegraphics[width=.45\linewidth]{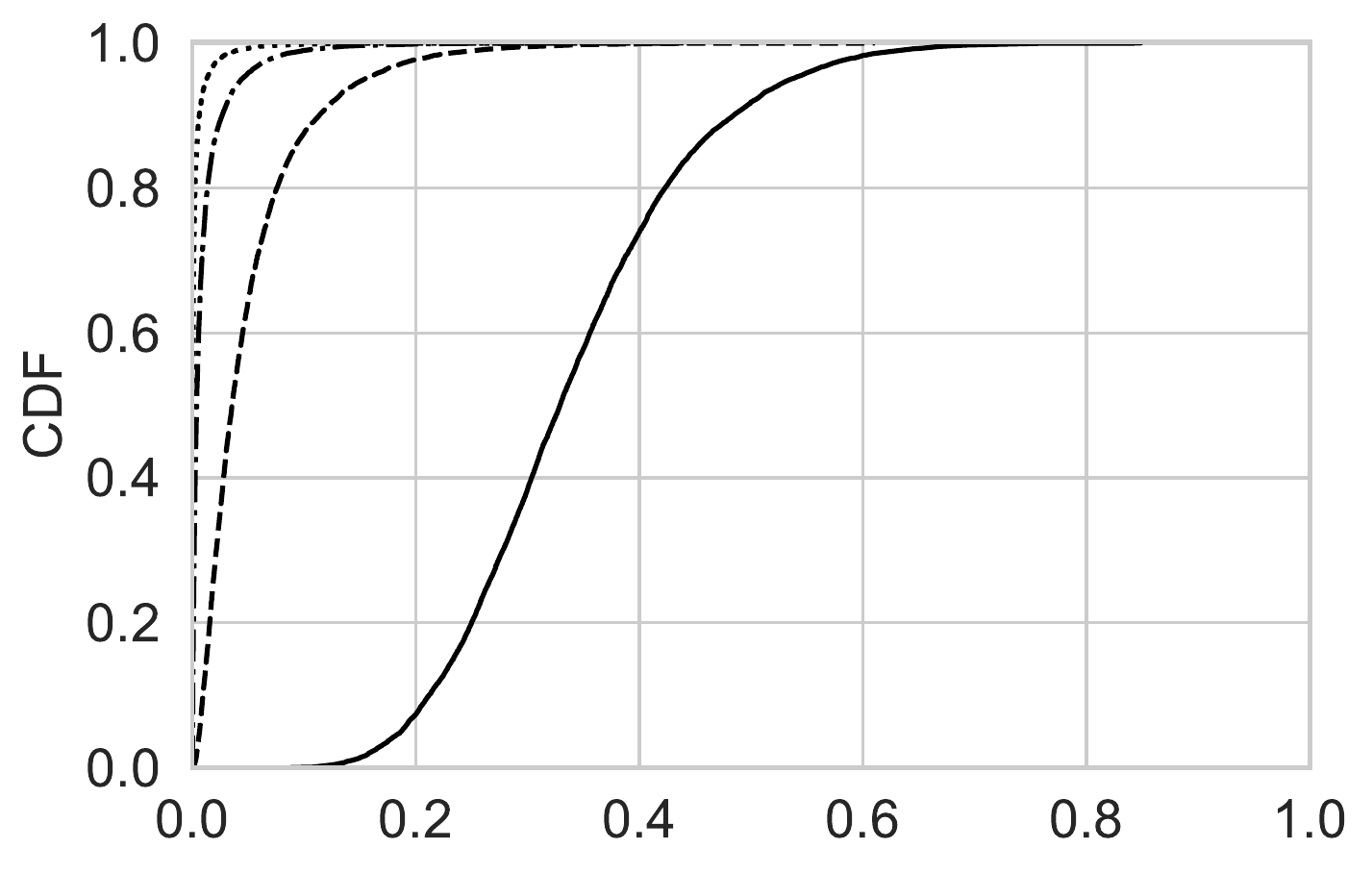}
         \par\medskip
         \includegraphics[width=.45\linewidth]{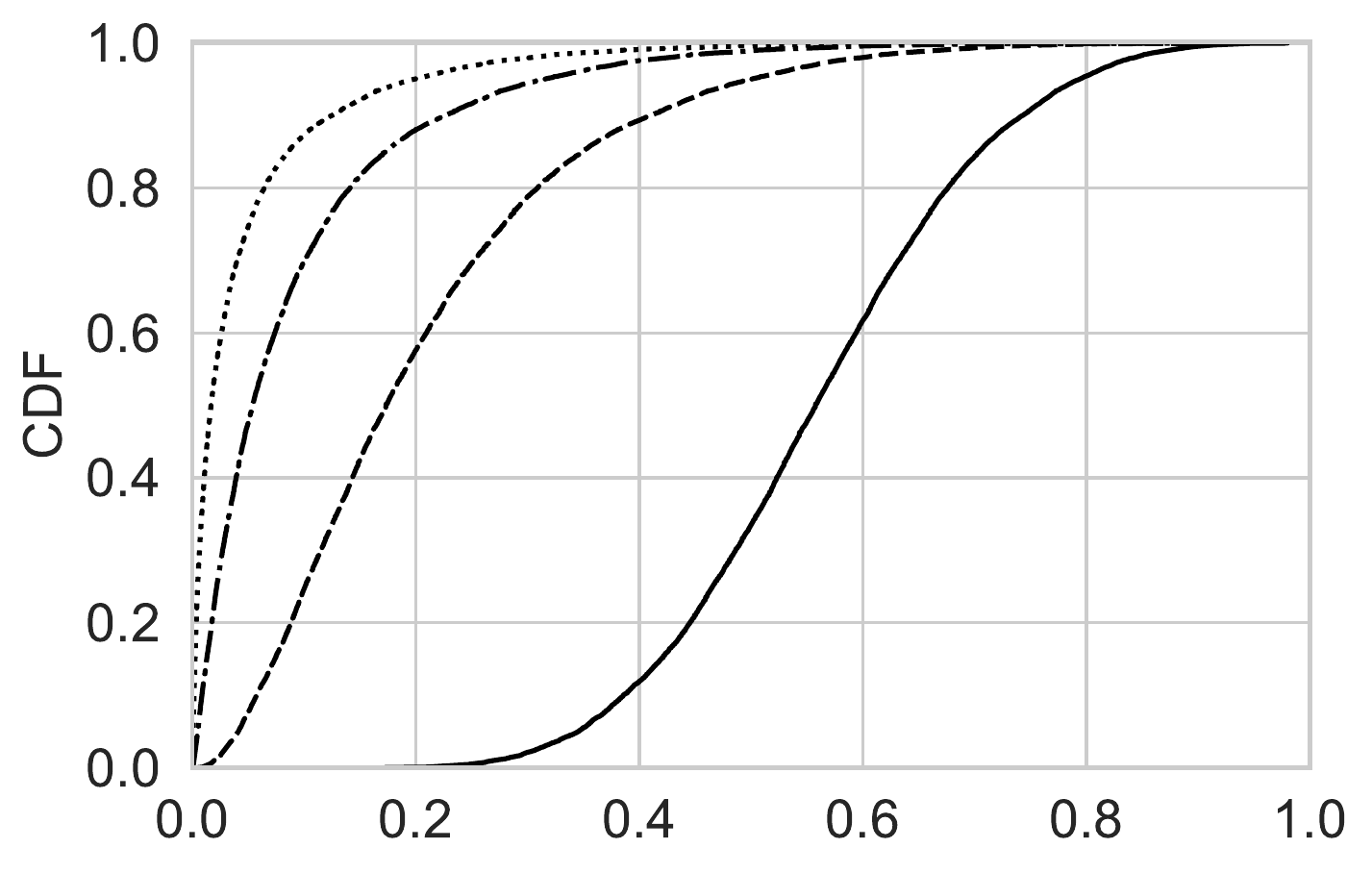}
    }
     \subfigure[$\alpha=0.3$ with $L=10$ (left) and $L=20$ (right)]{
         \includegraphics[width=.45\linewidth]{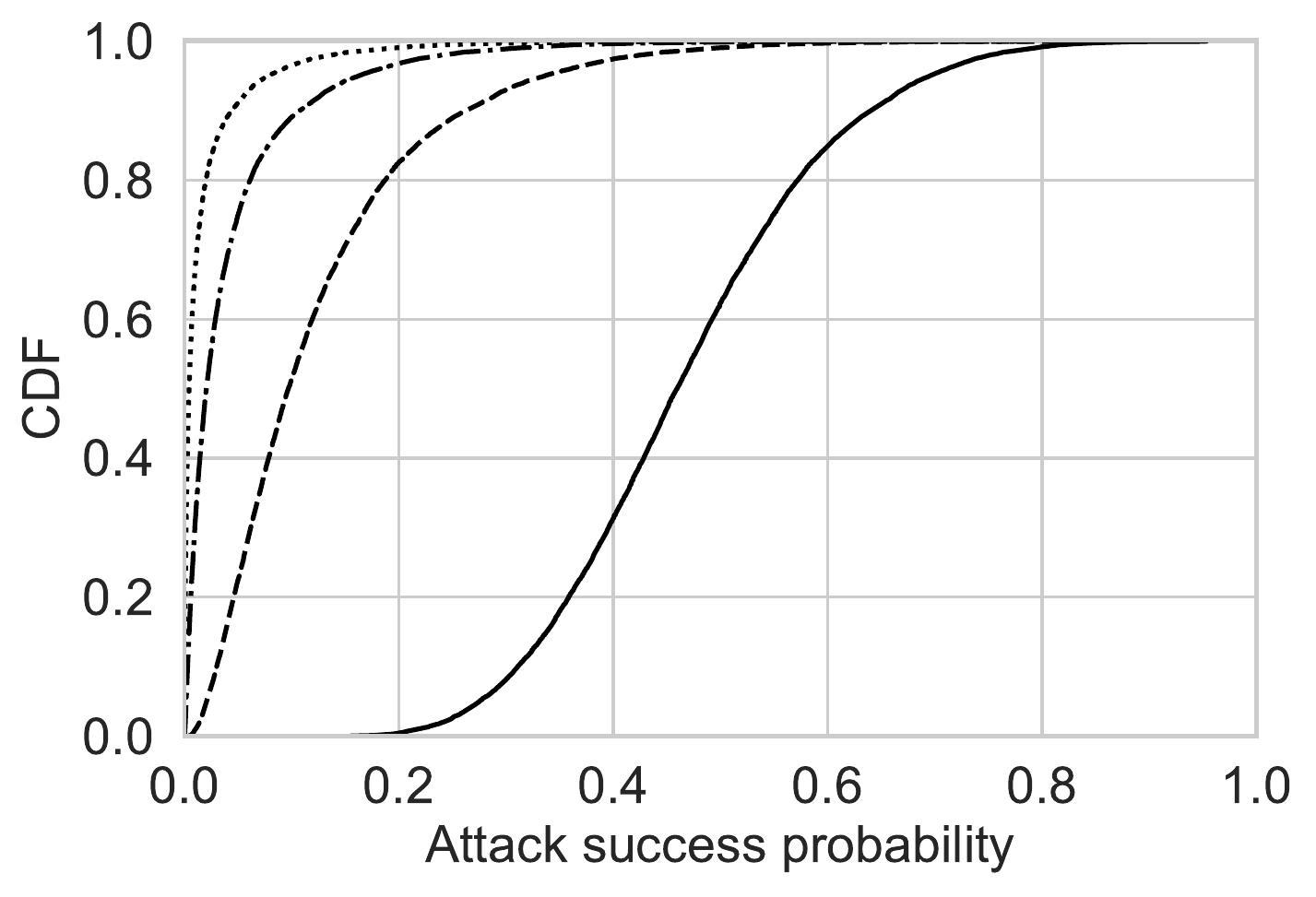}
         \par\medskip
         \includegraphics[width=.45\linewidth]{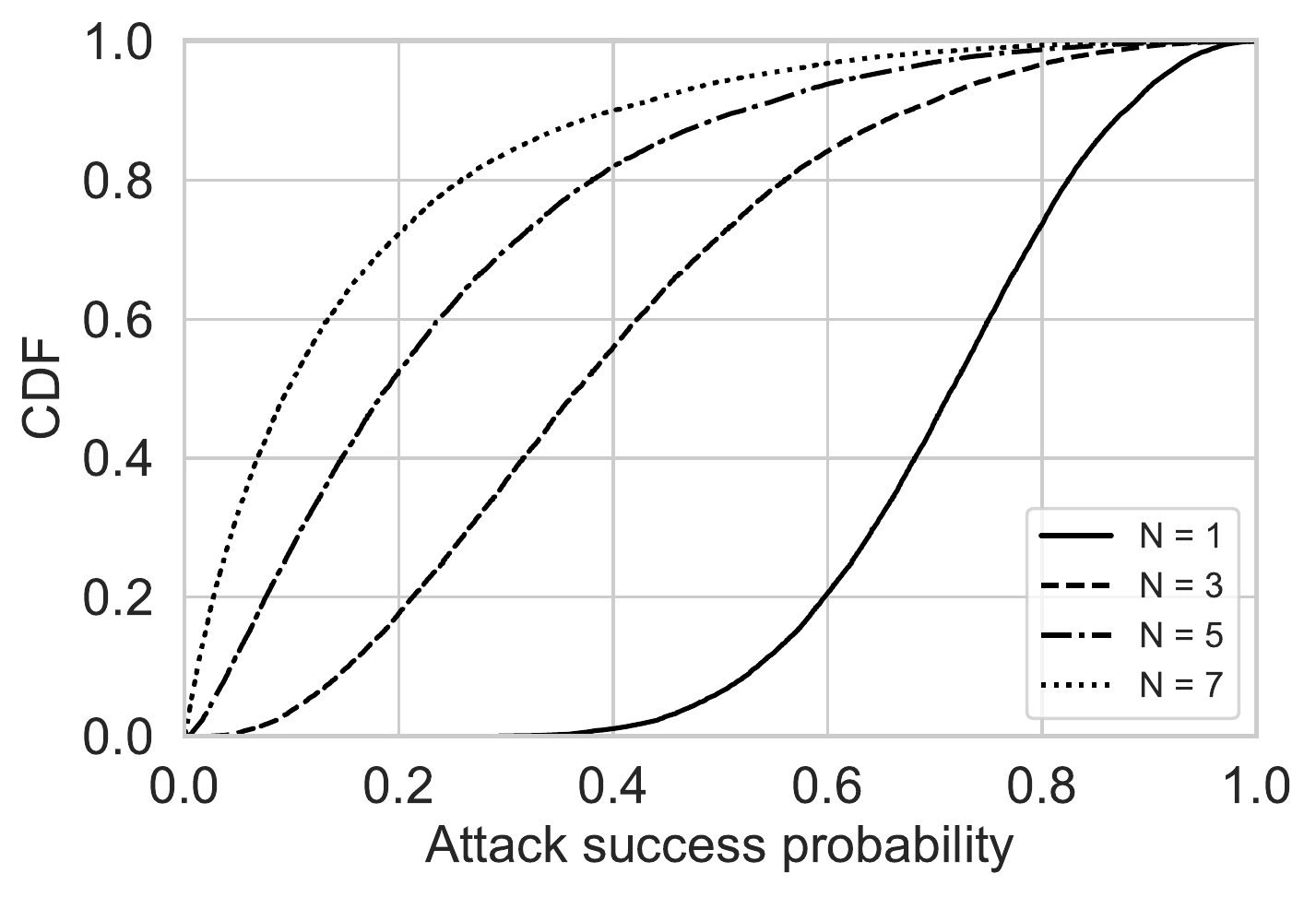}
    }
     \caption{Cumulative distribution functions (CDFs) of $P_{attack}(L, B, M, N)$ with $L$ = \{10, 20\} and $N$ = \{1, 3, 5, 7\} for 2,738 smart contracts. 
     }
     \label{fig:5_distortive_150}
\end{figure}

Figure~\ref{fig:5_distortive_150} shows the cumulative distribution functions (CDFs) when computing $P_{attack}(L, B, M, N)$ with $\alpha$ = \{0.2, 0.3\}, $L$ = \{10, 20\}, and $N$ = \{1, 3, 5, 7\} with the selected 11,646 
smart contracts 
for fair comparison (recall that the number of watermarkable contracts may be different depending on $L$ and $N$). 

Overall, choosing either a smaller $L$ or a larger $N$ raises the bar by requiring a higher execution cost for successful attacks, confirming our intuition on the resiliency of {\sys} against distortion attacks. With $L=10$, $N=7$, and $\alpha=0.2$, only $0.7\%$ of the smart contracts would be thwarted with the attack success probability of 0.05 or higher. Even when $\alpha$ increases 0.3 with the same configuration, only $8.9\%$ would be thwarted, being still effective. On the contrary, with $L=20$, $N=7$, and $\alpha=0.2$, the attack success probability significantly increases; $25.3\%$ would be thwarted with the attack success probability of 0.05 or higher. Based on these results, we advise not to utilize $L=20$ despite its superiority in credibility (Section~\ref{ss:Credibility}). Likewise, a single watermark (when $N$ = 1) would be ineffective against distortion attacks, indicating that multiple watermarks are recommended for \sys. It is noted that an adversary cannot increase $\alpha$ unreasonably because generated smart contracts with a large $\alpha$ are not competitive at all against the original smart contract in terms of execution cost. In a competitive DApp market, it is typical that a user avoids choosing a DApp that consumes an additional gas cost if there is an alternative to provide same/similar features. Likewise, Ethereum users are no difference~\cite{disc_usersensitive}.

\section{Discussions and Limitations} 

This section covers in-depth discussions and limitations of our approach, and future research.

\PP{Reversing EVM Bytecode}
EVM is a Turing complete virtual machine
based on the stack, which does not
follow the Von Neumann architecture
(\eg, no registers).
Besides, Solidity even complicates
bytecodes by introducing built-in functions and applying various optimizations.
In this regard, reverse engineering of an EVM bytecode and understanding underlying semantics (\ie, decompilation) are non-trivial~\cite{reversing-bytecode1, reversing-bytecode2, reversing-bytecode3}. This would make distortion attacks without source code quite challenging on a smart contract.

\PP{EVM Bytecode Diversification}
Having the identical source code, a runtime EVM bytecode may be diversified with the rapid evolution of the Solidity compiler~\cite{liu2019enabling}. However, the current design of {\sys} targets bytecode generation with the same compiler version. We conduct an additional experiment to confirm the robustness of {\sys} against different compiler versions (from 0.4.0 to 0.8.17). Our empirical results demonstrate {\WC{1}}~successful watermark detection with a minor version difference; of the total 5,028 contracts, only 6\% was undetected (L=15, N=3), and {\WC{2}}~cross-compilation failure with a different major version mostly due to unsupported syntaxes; {\sys} may be extended to support multiple compiler versions by creating an individual WRO per version, which we leave part of our future work. As a final note, recall that WROs are added to off-chain storage (rather than in a smart contract itself) without incurring an additional gas cost.

\PP{Threats to Validity}
The threats to the validity of this work mainly come from two aspects. 
A possible threat is whether we used representative smart contracts for evaluation. Even though we collected 4,112,336 smart contracts from all Ethereum blocks,
our experiments include only 27,824 smart contracts based on the DBSCAN clustering results for distinct
smart contracts. 
Before clustering, we exclude 7,422 (4.17\%) small-sized smart contracts because those contracts do not have a sufficient number of CFG blocks for embedding watermarks.
However, it would not be problematic with considerably complex business logic in most cases, which increases the size (and the number of blocks accordingly) of a smart contract.
Another threat to validity is the generalizability of \sys. Because the current implementation of \sys relies on EtherSolve~\cite{4_EtherSolve} to generate CFGs from smart contracts, our evaluation may not be applicable with a different CFG generator. Supporting additional CFG generators is part of our future work.

\section{Related Work}
\label{s:relwk}

\PP{Software Copyright and Smart Contracts} 
Protecting a software copyright often helps 
to maintain productivity and motivation of software development. 
Vast studies~\cite{peace2003software, samuelson2016functionality,
reavis1991software} have been conducted on the methods to detect 
and/or prevent a software piracy. 
Lately, the rise in popularity of 
smart contracts with the blockchain technology necessitates 
a new suitable means to safeguarding an ownership  \cite{savelyev2018copyright, bodo2018blockchain}.

\PP{Smart Contract Code Reuse} 
Recent studies repeatedly show that reusing code 
in a smart contract is quite 
prevalent~\cite{jia2020similar, kondo2020code,
kiffer2018analyzing, chen2021understanding, 
pierro2021analysis}.
Chen et al.~\cite{chen2021understanding} reveal
that 91.1\% out of 52,951 smart contract projects 
gathered by Etherscan~\cite{6_Etherscan} 
(before August 2019) contain 
one or more subcontracts from others.
According to the analysis by Pierro et 
al.~\cite{pierro2021analysis}, a wide adoption
of code reuse in a smart contract arises from 
the developers' desire for building
successful Ethereum DApps and/or the lack of
a well-integrated development tool.
While code reusability aids the quick and effortless development
of a smart contract, a security bug may bring
about an unwelcome outcome~\cite{he2020characterizing}
as seen in the past incidents~\cite{Popper2016Online,
Browne2017Online, Dale2021Online, M2018Online}.
In the meantime, EClone~\cite{liu2019enabling} detects
a replication of a smart contract based on its birthmark.
Note that \sys is designed for embedding and verifying a watermark while EClone aims to measure the similarity between smart contracts without considering security.

\PP{Software Watermarking Schemes} 
A wide spectrum of software watermarking 
techniques~\cite{Peter1994, Keith1994, davidson1996method, 
kang2021softmark, collberg1999software, 
jiang2009software, qu1998analysis, 
balachandran2014function, qu1999hiding, monden2000practical, lu2014ropsteg,
collberg2003sandmark, zuo2010zero, rovcek2021zero,
wang2019ternary, ma2015software}
have been proposed to protect a software copyright.
One of simple but efficient approaches leverages
code reordering~\cite{davidson1996method, kang2021softmark}
at the level of a basic block~\cite{davidson1996method}
or a function~\cite{kang2021softmark}, which inserts
a watermark by mapping it into the order of code.
Another well-studied direction for software watermarking 
utilizes a graph theory~\cite{collberg1999software, jiang2009software, qu1998analysis, qu1999hiding} such
as a graph coloring problem~\cite{qu1998analysis}.
Collberg et al.~\cite{collberg1999software} store
a graph structure on the heap at runtime.
Meanwhile, an obfuscation scheme has been widely
adopted~\cite{balachandran2014function, monden2000practical, lu2014ropsteg, collberg2003sandmark} in the field of software
watermarking, which includes
inserting dummy methods and opaque
predicates~\cite{monden2000practical}, and
steganography~\cite{lu2014ropsteg}.
Besides, the idea of embedding a watermark into a target 
application without modification (\ie, zero 
watermarking~\cite{zuo2010zero, rovcek2021zero,
wang2019ternary})
has been introduced, however, its downside lies in
needing additional storage for bookkeeping.
Meanwhile, Ma et al.~\cite{ma2015software} introduce an 
return-oriented-programming (ROP) based watermarking scheme,
which inserts a well-crafted code to be triggered 
into a data region for verification afterward.
However, applying prior software watermarking schemes
to a smart contract is impractical  
due to its unique properties such as the restriction of code 
size (\eg, code relocation, obfuscation),
the absence of dynamic allocation (\eg, runtime operation),
and execution costs (\eg, dummy code insertion).
\sys proposes a distinct watermarking scheme
tailored to smart contracts
for the first time by addressing the above hindrances.

\section{Conclusion}
\label{conclusion}
Smart contracts fundamentally have different characteristics from conventional programs. The difference induces several restrictions on adopting existing software watermark techniques to smart contracts. 
In this work, we present \sys, a novel watermarking scheme on smart contracts.
Our empirical evaluation shows the practicality and effectiveness 
of \sys from both security and economic perspectives,
which is resistant to various attacks against a watermark at acceptable cost.

\section{Data and Source Code Availability} 
\label{availability}
To foster further watermarking research for smart contracts, we disclose all of our evaluation dataset to the public. 
We have opened the datasets on a preserved digital repository\footnote{https://doi.org/10.6084/m9.figshare.21966875.v1} and source code\footnote{https://github.com/SKKU-SecLab/SmartMark.git}, so that anyone can reproduce our work.

\section*{Acknowledgements}
\label{s:ack}

We thank the anonymous reviewers for their constructive feedback.
It is noted that Yunhee Jang and Chanjong Lee are equally contributed as the second authors.
This work was partly supported by 
Institute of Information \& communications 
Technology Planning \& Evaluation (IITP) 
grant funded by the Korea government (MSIT) 
(No. 2022-0-01199; Graduate School of Convergence
Security (Sungkyunkwan university), No. 2022-0-00495; On-Device Voice Phishing Call Detection),
No. 2022-0-00688; AI Platform to Fully Adapt and Reflect Privacy-Policy Changes),
and the Basic Science Research Program through NRF
grant funded by the Ministry of Education
of the Government of South Korea 
(No. NRF-2022R1F1A1074373).
Any opinions, findings, and conclusions or 
recommendations expressed in
this material are those of the authors and 
do not necessarily reflect
the views of the sponsor.

\bibliographystyle{IEEEtran}
\bibliography{mybibliography}

\begin{thebibliography}{10}
\providecommand{\url}[1]{#1}
\csname url@samestyle\endcsname
\providecommand{\newblock}{\relax}
\providecommand{\bibinfo}[2]{#2}
\providecommand{\BIBentrySTDinterwordspacing}{\spaceskip=0pt\relax}
\providecommand{\BIBentryALTinterwordstretchfactor}{4}
\providecommand{\BIBentryALTinterwordspacing}{\spaceskip=\fontdimen2\font plus
\BIBentryALTinterwordstretchfactor\fontdimen3\font minus
  \fontdimen4\font\relax}
\providecommand{\BIBforeignlanguage}[2]{{%
\expandafter\ifx\csname l@#1\endcsname\relax
\typeout{** WARNING: IEEEtran.bst: No hyphenation pattern has been}%
\typeout{** loaded for the language `#1'. Using the pattern for}%
\typeout{** the default language instead.}%
\else
\language=\csname l@#1\endcsname
\fi
#2}}
\providecommand{\BIBdecl}{\relax}
\BIBdecl

\bibitem{he2020characterizing}
N.~He, L.~Wu, H.~Wang, Y.~Guo, and X.~Jiang, ``{Characterizing Code Clones in
  the Ethereum Smart Contract Ecosystem},'' in \emph{Proceedings of the 24th
  International Conference on Financial Cryptography and Data Security (FC)},
  2020, pp. 654--675.

\bibitem{Zhou18:Smartcontract}
Y.~Zhou, D.~Kumar, S.~Bakshi, J.~Mason, A.~Miller, and M.~Bailey, ``{Erays:
  Reverse Engineering Ethereum's Opaque Smart Contracts},'' in
  \emph{Proceedings of the 27th USENIX Security Symposium (Security)}, 2018,
  pp. 1371--1385.

\bibitem{brent18:Smartcontract}
L.~Brent, A.~Jurisevic, M.~Kong, E.~Liu, F.~Gauthier, V.~Gramoli, R.~Holz, and
  B.~Scholz, ``{Vandal: A Scalable Security Analysis Framework for Smart
  Contracts},'' \emph{arXiv preprint arXiv:1809.03981}, 2018.

\bibitem{Grech19:Smartcontract}
N.~Grech, L.~Brent, B.~Scholz, and Y.~Smaragdakis, ``{Gigahorse: Thorough,
  Declarative Decompilation of Smart Contracts},'' in \emph{Proceedings of the
  41st IEEE/ACM International Conference on Software Engineering (ICSE)}, 2019,
  pp. 1176--1186.

\bibitem{pierro2021analysis}
G.~A. Pierro and R.~Tonelli, ``{Analysis of Source Code Duplication in Ethereum
  Smart Contracts},'' in \emph{Proceedings of the 28th IEEE International
  Conference on Software Analysis, Evolution and Reengineering (SANER)}, 2021,
  pp. 701--707.

\bibitem{chen2021understanding}
X.~Chen, P.~Liao, Y.~Zhang, Y.~Huang, and Z.~Zheng, ``{Understanding Code Reuse
  in Smart Contracts},'' in \emph{Proceedings of the 28th IEEE International
  Conference on Software Analysis, Evolution and Reengineering (SANER)}, 2021,
  pp. 470--479.

\bibitem{collberg1999software}
C.~Collberg and C.~Thomborson, ``{Software Watermarking: Models and Dynamic
  Embeddings},'' in \emph{Proceedings of the 26th ACM SIGPLAN-SIGACT Symposium
  on Principles of Programming Languages (POPL)}, 1999, pp. 311--324.

\bibitem{Peter1994}
P.~R. Samson, ``{Apparatus and Method for Serializing and Validating Copies of
  Computer Software},'' \url{http://www.google.com/patents/US5287408A}, 1994.

\bibitem{Keith1994}
K.~Holmes, ``{Computer Software Protection},''
  \url{http://www.google.com/patents/US5287407}, 1994.

\bibitem{davidson1996method}
R.~I. Davidson and N.~Myhrvold, ``{Method and System for Generating and
  Auditing a Signature for a Computer Program},''
  \url{https://patents.google.com/patent/US5559884A}, 1996.

\bibitem{kang2021softmark}
H.~Kang, Y.~Kwon, S.~Lee, and H.~Koo, ``{SoftMark: Software Watermarking via a
  Binary Function Relocation},'' in \emph{Proceedings of the 37th Annual
  Computer Security Applications Conference (ACSAC)}, 2021, pp. 169--181.

\bibitem{jiang2009software}
Z.~Jiang, R.~Zhong, and B.~Zheng, ``{A Software Watermarking Method Based on
  Public-key Cryptography and Graph Coloring},'' in \emph{Proceedings of the
  3rd International Conference on Genetic and Evolutionary Computing (ICGEC)},
  2009, pp. 433--437.

\bibitem{qu1998analysis}
G.~Qu and M.~Potkonjak, ``{Analysis of Watermarking Techniques for Graph
  Coloring Problem},'' in \emph{Proceedings of the 1998 IEEE/ACM international
  Conference on Computer-Aided Design (ICCAD)}, 1998, pp. 190--193.

\bibitem{qu1999hiding}
------, ``{Hiding Signatures in Graph Coloring Solutions},'' in
  \emph{Proceedings of the 3rd Workshop on Information Hiding (IH)}, 1999, pp.
  348--367.

\bibitem{balachandran2014function}
V.~Balachandran, N.~W. Keong, and S.~Emmanuel, ``{Function Level Control Flow
  Obfuscation for Software Security},'' in \emph{Proceedings of the 8th
  International Conference on Complex, Intelligent and Software Intensive
  Systems (CISIS)}, 2014, pp. 133--140.

\bibitem{monden2000practical}
A.~Monden, H.~Iida, K.~Matsumoto, K.~Inoue, and K.~Torii, ``{A Practical Method
  for Watermarking Java Programs},'' in \emph{Proceedings of the 24th Annual
  International Computer Software and Applications Conference (COMPSAC)}, 2000,
  pp. 191--197.

\bibitem{lu2014ropsteg}
K.~Lu, S.~Xiong, and D.~Gao, ``{Ropsteg: Program Steganography with Return
  Oriented Programming},'' in \emph{Proceedings of the 4th ACM Conference on
  Data and Application Security and Privacy (CODASPY)}, 2014, pp. 265--272.

\bibitem{collberg2003sandmark}
C.~Collberg, G.~Myles, and A.~Huntwork, ``{Sandmark-A Tool for Software
  Protection Research},'' \emph{IEEE Security \& Privacy (S\&P)}, pp. 40--49,
  2003.

\bibitem{zuo2010zero}
J.~Zuo and D.~Cui, ``Zero-watermark resistant to mp3 compression,''
  \emph{Advanced Materials Research}, pp. 254--259, 2010.

\bibitem{rovcek2021zero}
A.~Ro{\v{c}}ek, M.~Javorn{\'\i}k, K.~Slav{\'\i}{\v{c}}ek, and O.~Dost{\'a}l,
  ``{Zero Watermarking: Critical Analysis of its Role in Current Medical
  Imaging},'' \emph{Journal of Digital Imaging}, pp. 204--211, 2021.

\bibitem{wang2019ternary}
C.~Wang, X.~Wang, Z.~Xia, and C.~Zhang, ``{Ternary Radial Harmonic Fourier
  Moments Based Robust Stereo Image Zero-watermarking Algorithm},''
  \emph{Information Sciences}, pp. 109--120, 2019.

\bibitem{ma2015software}
H.~Ma, K.~Lu, X.~Ma, H.~Zhang, C.~Jia, and D.~Gao, ``{Software Watermarking
  Using Return-oriented Programming},'' in \emph{Proceedings of the 10th ACM
  Symposium on Information, Computer and Communications Security (CCS)}, 2015,
  pp. 369--380.

\bibitem{collberg2004dynamic}
C.~Collberg, E.~Carter, S.~Debray, A.~Huntwork, J.~Kececioglu, C.~Linn, and
  M.~Stepp, ``{Dynamic Path-based Software Watermarking},'' in
  \emph{Proceedings of the 25th ACM SIGPLAN Conference on Programming Language
  Design and Implementation (PLDI)}, 2004, pp. 107--118.

\bibitem{Zhang20:Smartcontract}
M.~Zhang, P.~Zhang, X.~Luo, and F.~Xiao, ``{Source Code Obfuscation for Smart
  Contracts},'' in \emph{Proceedings of the 27th Asia-Pacific Software
  Engineering Conference (APSEC)}, 2020, pp. 513--514.

\bibitem{Ethereum2022Online}
Solidity, ``{Solidity — Solidity 0.8.13 Documentation},''
  \url{https://docs.soliditylang.org/en/v0.8.13/}, 2022.

\bibitem{Yan20:Smartcontract}
W.~Yan, J.~Gao, Z.~Wu, Y.~Li, Z.~Guan, Q.~Li, and Z.~Chen, ``{Eshield: Protect
  Smart Contracts against Reverse Engineering},'' in \emph{Proceedings of the
  29th ACM SIGSOFT International Symposium on Software Testing and Analysis
  (ISSTA)}, 2020, pp. 553--556.

\bibitem{szabo_formalizing_1997}
N.~Szabo, ``{Formalizing and Securing Relationships on Public Networks},''
  \emph{First Monday}, 1997.

\bibitem{wood2014ethereum}
G.~Wood \emph{et~al.}, ``{Ethereum: A Secure Decentralised Generalised
  Transaction Ledger},'' \emph{Ethereum Project Yellow Paper}, pp. 1--32, 2014.

\bibitem{Ethereum2021Online}
Solidity, ``{Layout of a Solidity Source File}, year = {2022},''
  \url{https://docs.soliditylang.org/en/v0.8.11/layout-of-source-files.html}.

\bibitem{wackerow2021Online}
Ethereum\;Foundation, ``{Opcodes for the EVM},''
  \url{https://ethereum.org/en/developers/docs/evm/opcodes/}, 2021.

\bibitem{2_Atzei}
N.~Atzei, M.~Bartoletti, and T.~Cimoli, ``{A Survey of Attacks on Ethereum
  Smart Contracts (SoK)},'' in \emph{Proceedings of the 6th International
  Conference on Principles of Security and Trust (POST)}, 2017, p. 164–186.

\bibitem{myles2005evaluation}
G.~Myles, C.~Collberg, Z.~Heidepriem, and A.~Navabi, ``{The Evaluation of Two
  Software Watermarking Algorithms},'' \emph{Software: Practice and
  Experience}, pp. 923--938, 2005.

\bibitem{zeng2010robust}
Y.~Zeng, F.~Liu, X.~Luo, and C.~Yang, ``{Robust Software Watermarking Scheme
  Based on Obfuscated Interpretation},'' in \emph{Proceedings of the 2nd
  International Conference on Multimedia Information Networking and Security
  (MINES)}, 2010, pp. 671--675.

\bibitem{dalla2017software}
M.~Dalla~Preda and M.~Pasqua, ``{Software Watermarking: A Semantics-based
  Approach},'' \emph{Electronic Notes in Theoretical Computer Science}, pp.
  71--85, 2017.

\bibitem{dey2019software}
A.~Dey, S.~Bhattacharya, and N.~Chaki, ``{Software Watermarking: Progress and
  Challenges},'' \emph{INAE Letters}, pp. 65--75, 2019.

\bibitem{jdourlens2020Online}
Ethereum\;Foundation, ``{Understanding the ERC-20 Token Smart Contract},''
  \url{https://ethereum.org/en/developers/tutorials/understand-the-erc-20-token-smart-contract/},
  2020.

\bibitem{4_EtherSolve}
F.~Contro, M.~Crosara, M.~Ceccato, and M.~Dalla~Preda, ``{EtherSolve: Computing
  an Accurate Control-flow Graph from Ethereum Bytecode},'' in
  \emph{Proceedings of the 29th IEEE/ACM International Conference on Program
  Comprehension (ICPC)}, 2021, pp. 127--137.

\bibitem{ethereum-standard}
Ethereum\;Foundation, ``{Ethereum-Development Standards},''
  \url{https://ethereum.org/en/developers/docs/standards/}, 2021.

\bibitem{6_clustering}
M.~Ester, H.-P. Kriegel, J.~Sander, and X.~Xu, ``{A Density-Based Algorithm for
  Discovering Clusters in Large Spatial Databases with Noise},'' in
  \emph{Proceedings of the 2nd International Conference on Knowledge Discovery
  and Data Mining (KDD)}, 1996, pp. 226–--231.

\bibitem{eip170}
Ethereum\;Foundation, ``{EIP-170: Contract Code Size Limit},''
  \url{https://eips.ethereum.org/EIPS/eip-170}, 2022.

\bibitem{disc_usersensitive}
J.~Jordan, ``{Daily Activity Down 87\%: Is Tether Killing Ethereum Gaming?}''
  \url{https://dappradar.com/blog/daily-activity-down-87-is-tether-killing-ethereum-gaming},
  2020.

\bibitem{reversing-bytecode1}
RET2\;Systems, ``{Building up from the Ethereum Bytecode},''
  \url{https://blog.ret2.io/2018/05/16/practical-eth-decompilation/}, 2022.

\bibitem{reversing-bytecode2}
F.~Sakharov, ``Reversing evm bytecode with radare2,''
  \url{https://blog.positive.com/reversing-evm-bytecode-with-radare2-ab77247e5e53},
  2022.

\bibitem{reversing-bytecode3}
B.~Arvanaghi, ``{Reversing Ethereum Smart Contracts},''
  \url{https://arvanaghi.com/blog/reversing-ethereum-smart-contracts/}, 2022.

\bibitem{liu2019enabling}
H.~Liu, Z.~Yang, Y.~Jiang, W.~Zhao, and J.~Sun, ``{Enabling Clone Detection for
  Ethereum via Smart Contract Birthmarks},'' in \emph{Proceedings of the 27th
  IEEE/ACM International Conference on Program Comprehension (ICPC)}, 2019, pp.
  105--115.

\bibitem{peace2003software}
A.~G. Peace, D.~F. Galletta, and J.~Y. Thong, ``{Software Piracy in the
  Workplace: A Model and Empirical Test},'' \emph{Journal of Management
  Information Systems}, pp. 153--177, 2003.

\bibitem{samuelson2016functionality}
P.~Samuelson, ``{Functionality and Expression in Computer Programs: Refining
  the Tests for Software Copyright Infringement},'' \emph{Berkeley Tech. LJ},
  p. 1215, 2016.

\bibitem{reavis1991software}
K.~Reavis~Conner and R.~P. Rumelt, ``{Software Piracy: An Analysis of
  Protection Strategies},'' \emph{Management Science}, pp. 125--139, 1991.

\bibitem{savelyev2018copyright}
A.~Savelyev, ``{Copyright in the Blockchain Era: Promises and Challenges},''
  \emph{Computer Law \& Security Review}, pp. 550--561, 2018.

\bibitem{bodo2018blockchain}
B.~Bod{\'o}, D.~Gervais, and J.~P. Quintais, ``{Blockchain and Smart Contracts:
  The Missing Link in Copyright Licensing?}'' \emph{International Journal of
  Law and Information Technology}, pp. 311--336, 2018.

\bibitem{jia2020similar}
N.~Jia, Q.~Kong, and H.~Huang, ``{How Similar are Smart Contracts on the
  Ethereum?}'' in \emph{Proceedings of the 2nd International Conference on
  Blockchain and Trustworthy Systems (BlockSys)}, 2020, pp. 403--414.

\bibitem{kondo2020code}
M.~Kondo, G.~A. Oliva, Z.~M.~J. Jiang, A.~E. Hassan, and O.~Mizuno, ``{Code
  Cloning in Smart Contracts: a Case Study on Verified Contracts from the
  Ethereum Blockchain Platform},'' \emph{Empirical Software Engineering}, pp.
  4617--4675, 2020.

\bibitem{kiffer2018analyzing}
L.~Kiffer, D.~Levin, and A.~Mislove, ``{Analyzing Ethereum's Contract
  Topology},'' in \emph{Proceedings of the 18th Internet Measurement Conference
  (IMC)}, 2018, pp. 494--499.

\bibitem{6_Etherscan}
Etherscan, ``{Ethereum (ETH) Blockchain Explorer},''
  \url{https://etherscan.io/}, 2022.

\bibitem{Popper2016Online}
N.~Popper, ``{A Hacking of More Than \$50 Million Dashes Hopes in the World of
  Virtual Currency},''
  \url{https://www.nytimes.com/2016/06/18/business/dealbook/hacker-may-have-removed-more-than-50-million-from-experimental-cybercurrency-project.html},
  2016.

\bibitem{Browne2017Online}
R.~Browne, ``{`Accidental' Bug May Have Frozen \$280 Million Worth of Digital
  Coin Ether in a Cryptocurrency Wallet},''
  \url{https://www.cnbc.com/2017/11/08/accidental-bug-may-have-frozen-280-worth-of-ether-on-parity-wallet.html},
  2017.

\bibitem{Dale2021Online}
B.~Dale, ``{Victims of \$30M Parity Wallet Hack Offer Attacker \$60M
  ‘Bounty’},''
  \url{https://finance.yahoo.com/news/victims-30m-parity-wallet-hack-170550649.html},
  2021.

\bibitem{M2018Online}
T.~M, ``{Pigeoncoin – The Coin That Couldn’t Fly},''
  \url{https://bitcoin.co.uk/pigeoncoin-the-coin-that-couldnt-fly}, 2018.

\end{thebibliography}

\end{document}